\documentclass[floatfix,secnumarabic,amssymb,nobibnotes,nofootinbib,preprint,aps,pra, superscriptaddress,showpacs]{revtex4-2}
\usepackage{amsmath}
\usepackage{amsfonts}
\usepackage{amssymb}
\usepackage{graphicx}
\usepackage{units}
\usepackage{color}
\usepackage{xspace}
\usepackage[normalem]{ulem}
\usepackage{natbib}
\usepackage{textcomp}

\newcommand{\fig}[1]{Fig.~\ref{#1}}

\DeclareUnicodeCharacter{2009}{\,}
\DeclareUnicodeCharacter{2212}{-}

\begin{document}

\title{Pressure induced insulator-to-metal transition in few-layer FePS$_3$ at 1.5 GPa}

\author{Bidyut Mallick$^{\dag}$}
\affiliation{Department of Applied Science and Humanities, Galgotias College of Engineering and Technology, Knowledge Park-II, Greater Noida, Uttar Pradesh - 201310, India}
\author{Mainak Palit$^{\dag}$}
\affiliation{School of Physical Sciences, Indian Association for the Cultivation of Science, 2A \& 2B Raja S. C. Mullick Road, Jadavpur, Kolkata - 700032, India
}

\author{Rajkumar Jana$^{\dag}$}
\affiliation{School of Chemical Sciences, Indian Association for the Cultivation of Science, 2A \& 2B Raja S. C. Mullick Road, Jadavpur, Kolkata - 700032, India
}
\author{Soumik Das}
\affiliation{School of Physical Sciences, Indian Association for the Cultivation of Science, 2A \& 2B Raja S. C. Mullick Road, Jadavpur, Kolkata - 700032, India
}

\author{Anudeepa Ghosh}
\affiliation{School of Physical Sciences, Indian Association for the Cultivation of Science, 2A \& 2B Raja S. C. Mullick Road, Jadavpur, Kolkata - 700032, India
}

\author{Janaky Sunil}
\affiliation{Chemistry and Physics of Materials Unit, Jawaharlal Nehru Centre for Advanced Scientific Research, Jakkur P.O., Bangalore - 560064, India
}
\author{Sujan Maity}
\affiliation{School of Physical Sciences, Indian Association for the Cultivation of Science, 2A \& 2B Raja S. C. Mullick Road, Jadavpur, Kolkata - 700032, India
}

\author{Bikash Das}
\affiliation{School of Physical Sciences, Indian Association for the Cultivation of Science, 2A \& 2B Raja S. C. Mullick Road, Jadavpur, Kolkata - 700032, India
}
\author{Tanima Kundu}
\affiliation{School of Physical Sciences, Indian Association for the Cultivation of Science, 2A \& 2B Raja S. C. Mullick Road, Jadavpur, Kolkata - 700032, India
}

\author{
Chandrabhas Narayana}
\affiliation{Chemistry and Physics of Materials Unit, Jawaharlal Nehru Centre for Advanced Scientific Research, Jakkur P.O., Bangalore - 560064, India
}

\author{Ayan Datta}
\affiliation{School of Chemical Sciences, Indian Association for the Cultivation of Science, 2A \& 2B Raja S. C. Mullick Road, Jadavpur, Kolkata - 700032, India
}
\author{Subhadeep Datta}
\email{sspsdd@iacs.res.in}
\affiliation{School of Physical Sciences, Indian Association for the Cultivation of Science, 2A \& 2B Raja S. C. Mullick Road, Jadavpur, Kolkata - 700032, India
}

\def\thefootnote{}\footnotetext{$^\dagger$ These authors contributed equally to this work}\def\thefootnote{\arabic{footnote}}

\begin{abstract}

In two-dimensional (2D) van der Waals (vdW) layered materials the application of pressure often induces a giant lattice collapse, which can subsequently drive an associated Mott transition. Here, we investigate room-temperature layer-dependent insulator-metal transition (IMT) and probable spin-crossover (SCO) in vdW magnet, FePS$_3$, under high-pressure using micro-Raman scattering. Experimentally obtained spectra, in agreement with the computed Raman modes, indicate evidence of IMT of FePS$_3$ started with a thickness-dependent critical pressure ($P_c$) which reduces to 1.5 GPa in trilayer flakes compared to 10.8 GPa for the bulk counterpart. Using a phenomenological model, we argue that strong structural anisotropy in few-layer flakes enhances the in-plane strain under applied pressure and is, therefore, ultimately responsible for reducing the critical pressure for the IMT with decreasing layer numbers. Reduction of the critical pressure for phase transition in  vdW magnets to 1-2 GPa marks the possibility of using intercalated few-layers in the field-effect transistor device architecture, and thereby, avoiding the conventional use of the diamond anvil cell (DAC).

\end{abstract}

\maketitle

\section{Introduction}

Correlation between electronic and magnetic properties in quantum materials can be realized in spatially confined 2D flakes of vdW magnets under high pressure.  
Moreover, with varying layer thickness \textit{i.e.} reducing the flake thicknesses down to a few-atomic layer, electronic band structure and spin texture can be altered, as a result of which charge transport and magnetism can be controlled \cite{Mak2010, Wang2012, Tongay2012, Dai2014, Liu2016, Nayak2015}. Stacking of isolated monolayers and multilayers of dissimilar materials led to the finding of vdW heterostructures and superlattices with exotic behaviour resulting in practical electronic and spintronic devices \cite{Geim2013, Novoselov2016}. Recently, the discovery of 2D long-range magnetic ordering and strong coupling in the spin-valley degree of freedom in 2D semiconductors has generated considerable interest as these materials would be an ideal building blocks to design multistate spin logic  \cite{Wolf2001, Jungwirth2016}. Further, a 2D electronic system with quasi-one-dimensional triangular spin-lattice shows antiferromagnetic ordering at ambient pressure, but superconductivity under high pressure \cite{Hiroshi}. Even though theoretical studies have proposed a number of 2D magnets down to monolayer limit \cite{Ma2012, Lebegue2013, Sivadas2015, Zhuang2015, Zhang2015},  very few experimental reports concern 2D flakes \cite{Sandilands2010, Kuo2016, Du2016, Tian2016}. Thus, tunable spin-ordering in few-layer magnets in comparison to their bulk counterpart is important for exploring exotic phases in new device architecture. A substantial amount of work has been done on the magnetism of bulk 2D transition metal phosphorous trichalcogenide (MPX$_3$) materials \cite{Lee2016}. The magnetization of these materials varies with the $d$-electron configurations of the transition metal (M): Ising type ordering in FePS$_3$ \cite{Lee2016}, Heisenberg type ordering in MnPS$_3$ \cite{Sun2019} and XY type ordering in NiPS$_3$ \cite{Wildes2015}. Among these, FePS$_3$ has attracted particular interest as Ising type magnetic ordering in the bulk FePS$_3$, which is stable in air \cite{Du2016}, persists down to monolayer \cite{Lee2016}.

In bulk FePS$_3$ with monoclinic crystal structure (space group C2/m), each Fe$^{2+}$ ion is bonded to six S$^{2-}$ ions to form edge-sharing FeS$_{6}$ octahedra, while the Fe atoms form a honeycomb lattice. The S atoms are connected with two P atoms just above and below the Fe honeycomb plane which constitutes (P$_2$S$_6$)$^{4-}$ unit \cite{Lee2016}. These arrangements are stacked along the $c$ axis where each plane is bonded via very weak interlayer vdW interaction. The anisotropic magnetic behaviour of the system is governed by the competing direct Fe-Fe exchange and indirect Fe-S-Fe superexchange interactions within layers (0.31 meV/cell), as well as by interlayer exchange interactions (1.41 meV/cell) \cite{Lee2016}.    

Utilizing hydrostatic pressure as external stimuli and Raman scattering as an experimental probe to identify changes in phonon modes of a 2D magnet in few-layers may facilitate the detection of correlated structural change in a progressive manner with enhancement of $d$-$p$ metal-ligand charge transfer, and possible SCO \cite{Wang2018, Haines2018}. To the best of our knowledge, except the study on quantifying vdW interactions in the layered MoS$_2$ by measuring the valence band maximum (VBM) splitting under pressure \cite{pengchong}, pressure-driven phase transition in few-layer vdW magnets have not yet been systematically investigated. The existing reports do not discount the effect of the third dimension in high pressure studies on these 2D systems \cite{Wang2018, Wang2016, Ivan2021, Pawbake2022}. Consequently, the nature of volume collapse in finite-sized 2D materials, and the effect of the crystal field splitting with $d^{4-7}$ transition-metal-elements while reducing the sample thickness, and a possible concurrent IMT may add new scaling laws, originating from the dimensional effect, to the existing band theory of Mott insulators.

Herein, we methodically investigate room-temperature pressure-driven IMT in FePS$_3$ from the bulk ($\sim$ 100 layers) to nearly 2D limit ($\sim$ 3 layers) following characteristic phonon modes using micro-Raman scattering. Raman modes, compared with the first-principles density functional theory (DFT) calculations, reveal a thickness-dependent critical pressure ($P_c$) which reduces to 1.5 GPa in 3-layer flakes compared to 10.8 GPa for the bulk counterpart.We have adopted a phenomenological model where a macroscopic structural change due to the variation of the effective in-plane strain with layer numbers may result in a reduction of the critical pressure in few-layer samples. Under applied pressure, few-layer flakes experience enhanced in-plane strain due to strong structural anisotropy which stipulates the decreasing effect of thickness on the critical pressure for IMT. Practically, interfacial adhesion in stacked vdW heterostructure or molecules trapped between layers can be used to realise pressure as low as 1-2 GPa  \cite{Trap2016}. The resulting conformational or magnetic changes can be investigated without use of intricate DAC methodology. Most importantly, engineered flakes can be employed in the conventional three-terminal field-effect transistor (FET) device architecture.

\section{Experimental and Computational Details}

Bulk crystals of FePS$_3$ with 99.999 $\%$ purity from 2D Semiconductors (USA) were used in the present study. Three types of FePS$_3$ samples, namely thin layer FePS$_3$ flake, thick layer FePS$_3$ flake and bulk FePS$_3$, were investigated during the experiment. 
FePS$_3$ flakes were mechanically exfoliated from the bulk crystal using scotch tape. 
The exfoliated flakes were transferred onto the diamond anvil using a micromanipulator under an optical microscope such that the flakes were placed at the centre of the diamond culet (Fig. S1). 
A stainless steel gasket was used as a sample holder. The gasket was indented to a thickness of 60 $\mu$m and a hole of 180 $\mu$m diameter was drilled at the centre of indentation to create the sample chamber.  A few Ruby spheres were placed with the sample to determine the pressure inside the sample chamber during the experiment. The chamber was filled with methanol-ethanol mixed in 4:1 ratio which acted as a quasi-hydrostatic pressure transmitting medium (PTM). 
The flakes are not controlled in thickness and each flake contains areas with various thicknesses. The thickness of the target experimental area of a flake was estimated following previous experience as reported in the earlier studies \cite{Ghosh2021, Palit2021}. The experiment was done in a Diacell Bragg-Mini diamond anvil cell (DAC) with a culet diameter of 500 $\mu$m. 
Raman spectra were collected using a Horiba T64000 Raman set-up. The spectra were taken in backscattering geometry using a DPSS laser ($\lambda$ = 532 nm) as excitation source, and the beam was focused to a spot size of 1 $\mu$m by 50x long distance objective (NA = 0.50). A constant laser power of around 5 mW was used throughout the experiment. We kept the temperature of the surroundings fixed during the experiment. Any small temperature change could not affect our ruby spectra within the spectral resolution of the Raman spectrometer. Hence the pressure values are mentioned up to the first decimal places with ±0.1GPa error.  The consistency of Raman modes for a particular thickness was verified at different spots with similar contrast under an optical microscope. 

All the spin-polarized calculations were performed within the framework of density functional theory (DFT) using the plane-wave technique as implemented in the Vienna Ab Initio Simulation Package (VASP) \cite{Kresse1993}.
Further details of the computational studies are given in Supplementary Information(SI) \cite{SI}.

\section{Experimental and Computational Results}

High-pressure x-ray diffraction study on powder FePS$_3$ by Haines \textit{et al.} \cite{Haines2018} reported two structural phase transitions, the first transition pressure (PT1) at approximately 4 GPa and the second (PT2) at approximately 14 GPa.  At the first transition pressure, an alignment of the vdW planes was predicted, where Fe atoms directly come above one another, and likewise for the P atoms. At PT2 the study observed dramatic volume collapse in the out-of-plane direction.  The study also reported an insulator-to-metal transition at around 5 GPa in resistivity measurement on a single crystal. Similar structural evolution was observed by Jarvis \textit{et al.} in x-ray diffraction study on powder and single crystal FePS$_3$  \cite{Jarvis2023}.
Further, Wang \textit{et al.}'s resistivity measurements on single crystalline FePS$_3$ and x-ray emission spectroscopy study showed the insulator to metal transition and spin state transition of Fe$^{2+}$ ions at around 14 GPa, which was accompanied by a sudden volume collapse in high-pressure powder x-ray diffraction study \cite{Wang2018}. 
The present study reports the first high-pressure Raman measurements on mechanically exfoliated vdW layers, here FePS$_3$, under quasi-hydrostatic pressure in an attempt to tune the transition pressure (PT1 as suggested by Haines \textit{et al.} \cite{Haines2018}) for the alignment of vdW planes depending on the layer numbers. The experimental observations for three different layer thicknesses have been corroborated by computational calculations. 

\begin{figure}[ht]
\centerline{\includegraphics[scale=0.8, clip]{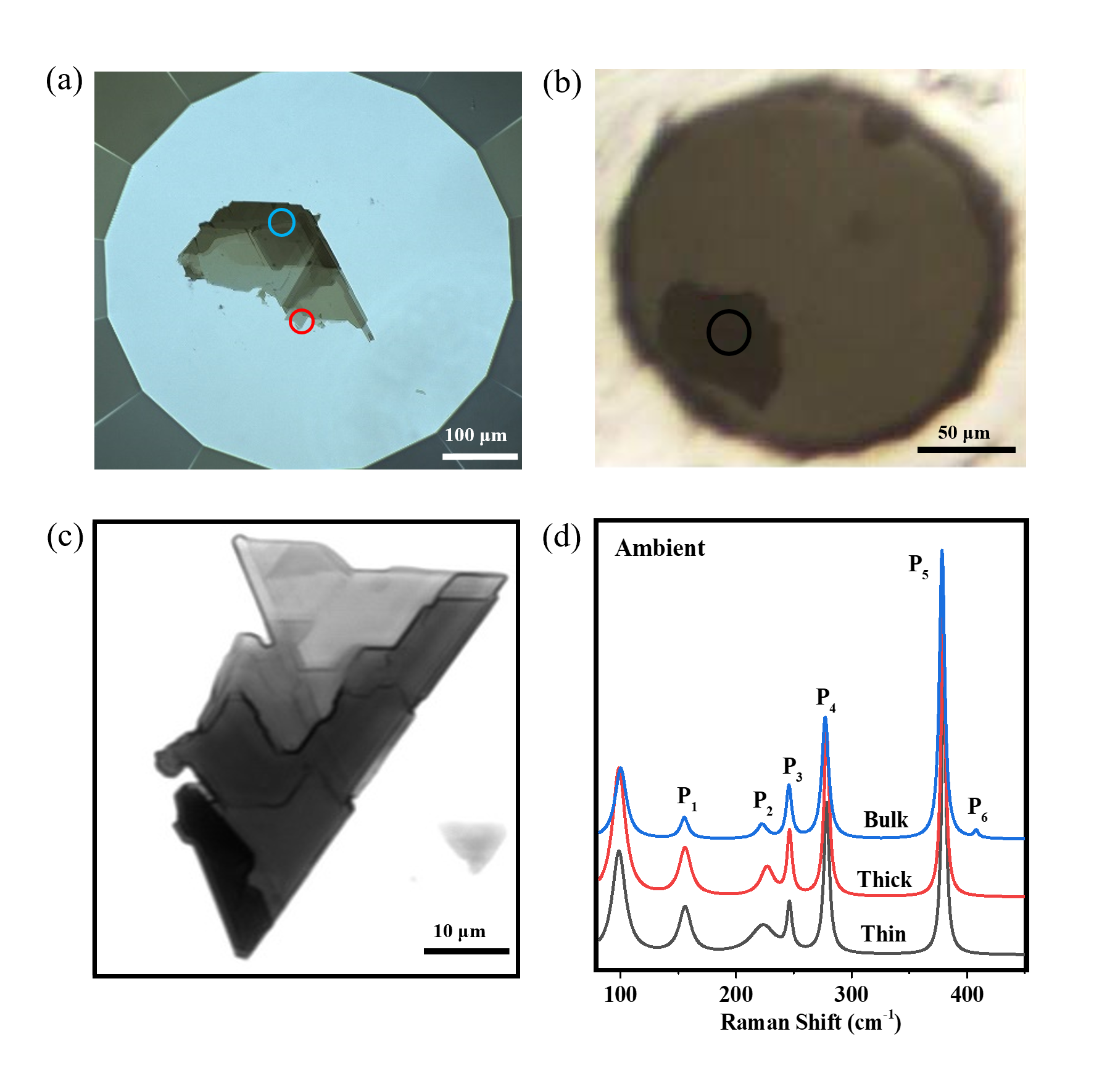}}
\caption{Raman spectroscopy of FePS$_3$ flake on diamond culet: (a) Micrograph of an exfoliated flake on the diamond culet: thin and thick areas are marked with red and blue circles respectively. (b) Transferred bulk sample inside the loaded DAC marked with a black circle. (c) Optical micrograph in transmission mode of a typical exfoliated flake showing the contrast variation with change in layer number. (d) Room temperature Raman spectra of the bulk, thick and thin layer FePS$_3$ on top of the diamond. Different phonon modes are labelled as P$_i  (i=1-6)$. A new weak mode P$_6$ is spotted at 407 cm$^{-1}$, which is not reported in the previous studies. P$_6$ is not clearly visible for thick and thin layers.
The broad Raman mode around 105 cm$^{-1}$ is a zone-folded phonon mode, as reported earlier \cite{Ghosh2021}.
\label{fig1}}
\end{figure}

Three types of FePS$_3$ samples were transferred on to the diamond culet: thin layer ($\sim$ 3 layers), thick layer ($\sim$ 30 layers) and bulk (more than 100 layers). Examples of the first two are shown in Fig.1(a). \fig{fig1}(b) depicts the image of the bulk sample (more than 100 layers) loaded inside the gasket hole in the DAC. In conclusion, substrate-free FePS$_3$ samples were directly loaded into the DAC to investigate the intrinsic properties of the two-dimensional systems, minimizing the effect of strain, substrate-material charge transfer or optical interference. 
The approximate layer thickness of the transferred flakes has been determined from the intensity ratio of the Raman modes at 277 and 378 cm$^{-1}$ \cite{Ghosh2021, Palit2021, SurfaceModGaIon2019}. This has been reconfirmed by an optical micrograph in transmission mode  (\fig{fig1}(c)) of the exfoliated samples on diamond culet where darker contrast indicates a higher number of layers.
\fig{fig1}(d) shows the Raman spectra of the thin layer, thick and bulk  FePS$_3$ at room temperature, which is similar to the previous reports \cite{Lee2016, Wang2016, Scagliotti1987}. At ambient condition,  FePS$_3$ adopts the C2/m symmetry group, which has 30 vibrational modes at the Brillouin zone centre: $\Gamma$ = 8A$_g$+6A$_u$+7B$_g$+9B$_u$.
Among these, the Raman active A$_g$ and B$_g$ modes are observed in the Raman scattering experiment.   
Previous Raman scattering data on FePS$_3$ as reported  by Wang \textit{et al.} \cite{Wang2018} identified the phonon symmetries of the observed modes  P$_1$, P$_2$ and P$_4$ as A$_g$/B$_g$ (156, 224 \& 277 cm$^{-1}$) and P$_3$ and P$_5$ as A$_g$ (246 and 378 cm$^{-1}$); these modes are due to molecule like vibration from the (P$_2$S$_6$)$^{4-}$ bipyramid structure. The low-frequency peaks at 98 cm$^{-1}$ and 156 cm$^{-1}$ are from the vibration of Fe atoms and the broadening of these modes is due to local fluctuation or disorder as suggested by Lee \textit{et al} \cite{Lee2016}. Another new weak peak detected for the bulk sample at 407 cm$^{-1}$ (P$_6$) was not reported in previous studies. Apart from the diamond anvil, FePS$_3$ with different layer thicknesses have been transferred and characterized on SiO$_2$/Si substrate, $p$-Si, hBN, graphene and diamond thin films. The micro-Raman spectrum in each case contains the characteristic Raman modes discussed in this report. The individual peaks have finite shifts for different substrates, but importantly, the relative positions of the specific modes remain unchanged. Since the current study investigates the evolution of the Raman modes with pressure, the relevance of such substrate effect may be discounted.

\begin{figure}[ht]
\centerline{\includegraphics[scale=0.5, clip]{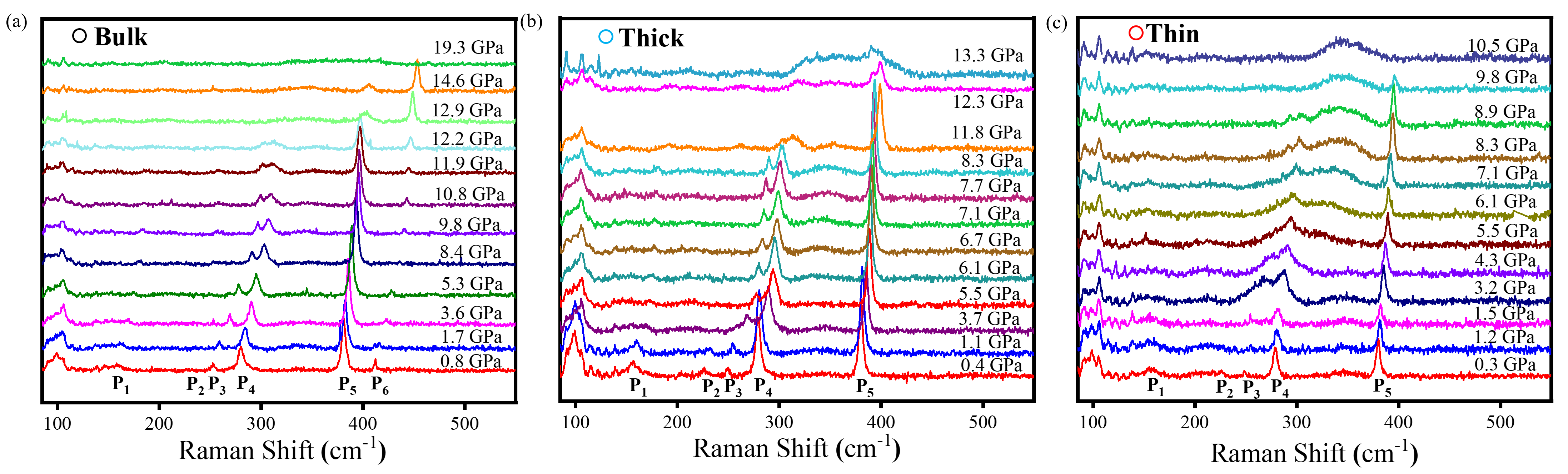}}
\caption{Pressure dependent Raman spectroscopy of layered FePS$_3$: Pressure responses of Raman modes for (a) bulk, (b) thick and (c) thin FePS$_3$ flakes were recorded in a diamond anvil cell (DAC). All the modes P$_i (i= 1-6) $ exhibit blue shift with increasing pressure. P$_1$ and P$_2$ are weak in intensity. Unlike the other peaks, with a change in pressure, the mode at 98 cm$^{-1}$ does not shift, rather a change in its shape is detected for all three samples. For bulk samples, P$_3$ and P$_4$ modes disappear by 12.9 GPa. The intensity of P$_5$ decreases and P$_6$ becomes more intense. For thick layers, P$_3$ and P$_4$ become very broad at 11.8 GPa and difficult to detect after that. P$_3$ and P$_4$ modes become broad and asymmetric above 1.5 GPa for thin sample. P$_3$, P$_4$ and P$_5$ disappear below 10 GPa, and a broad peak between 300 - 370 cm$^{-1}$ persists at high pressure.  
\label{fig2}}
\end{figure}

In the bulk FePS$_3$, \fig{fig2}(a), with increasing pressure, the Raman spectra show a hardening of P$_1$ to P$_6$ up to 12.2 GPa. Note that P$_2$ at 224 cm$^{-1}$ cannot be detected with considerable clarity.  By 12.9 GPa, P$_3$ and P$_4$ disappear. The intensity of P$_5$ decreases and P$_6$ becomes more intense. With the application of pressure, all the detectable Raman peaks shift to higher frequencies because of the strengthening of intra-atom interactions under hydrostatic pressure. Unlike the other peaks, the frequency shift of the mode at 98 cm$^{-1}$ with pressure has not been observed, rather a change in its shape is detected with the increasing pressure. The fact that all the modes seem to be broadening above 10 GPa, may be due to the hydrostatic limit of pressure transmitting medium \cite{PTM2009}. P$_3$ and P$_4$ overlap and become difficult to be recognized beyond 12.2 GPa.  
The pressure-induced line broadening for all the peaks, along with the complete loss of Raman intensity can be attributed to pressure-driven metallization of FePS$_3$. 
Also, the evolution of the peak at 98 cm$^{-1}$ under compression can be attributed to the gradual shift of the Fe atoms in the hexagon plane with pressure during the alignment of the vdW planes. The Raman modes for the bulk sample recover when the pressure is released (Fig. S2(a)).
Concerning thick layer FePS$_3$ flake, in \fig{fig2}(b),  all the characteristic Raman modes of FePS$_3$ can be identified, except the P$_6$ mode. Importantly, the Raman spectra exhibit the same trend as that observed in the bulk sample.  Similar to the bulk sample, the broadening of all the Raman modes is observed above 10 GPa. While the pressure is released all the modes recovered around 3.6 GPa (Fig. S2(b)).	

Unlike bulk or thick sample, for the thin-layered FePS$_3$, \fig{fig2}(c), P$_1$ and P$_2$ are very weak in intensity whereas the evolution of P$_3$, P$_4$ and P$_5$  characteristic modes can be observed with clarity. A similar trend in peak shift has been observed with increasing pressure, but noticeably, P$_3$ and P$_4$ modes become broad and asymmetric above 2 GPa. The asymmetric line shape of the Raman spectra at the onset of metallization has also been observed for several other magnetic compounds \cite{Marrocchelli2007, Her2006, Baldini2011}. 
Hence, the metallization of thin layer FePS$_3$ may initiate below 2 GPa. P$_3$, P$_4$ and P$_5$ disappear by 10 GPa, suggesting complete metallization of thin layer FePS$_3$. A broad peak persists between  300 cm$^{-1}$ and 370 cm$^{-1}$ at high pressure. The P$_3$, P$_4$ and P$_5$ modes recover when the pressure is released, Fig. S2(c).  
The decrease of metallization pressure with decreasing thickness suggests interlayer interactions of FePS$_3$ opposing IMT.  The P$_6$ mode is prominent above 12 GPa in the bulk sample and is not detected in thin and thick layered samples may be due to less number of vdW planes. 
 
\begin{figure}[ht]
\centerline{\includegraphics[scale=0.7, clip]{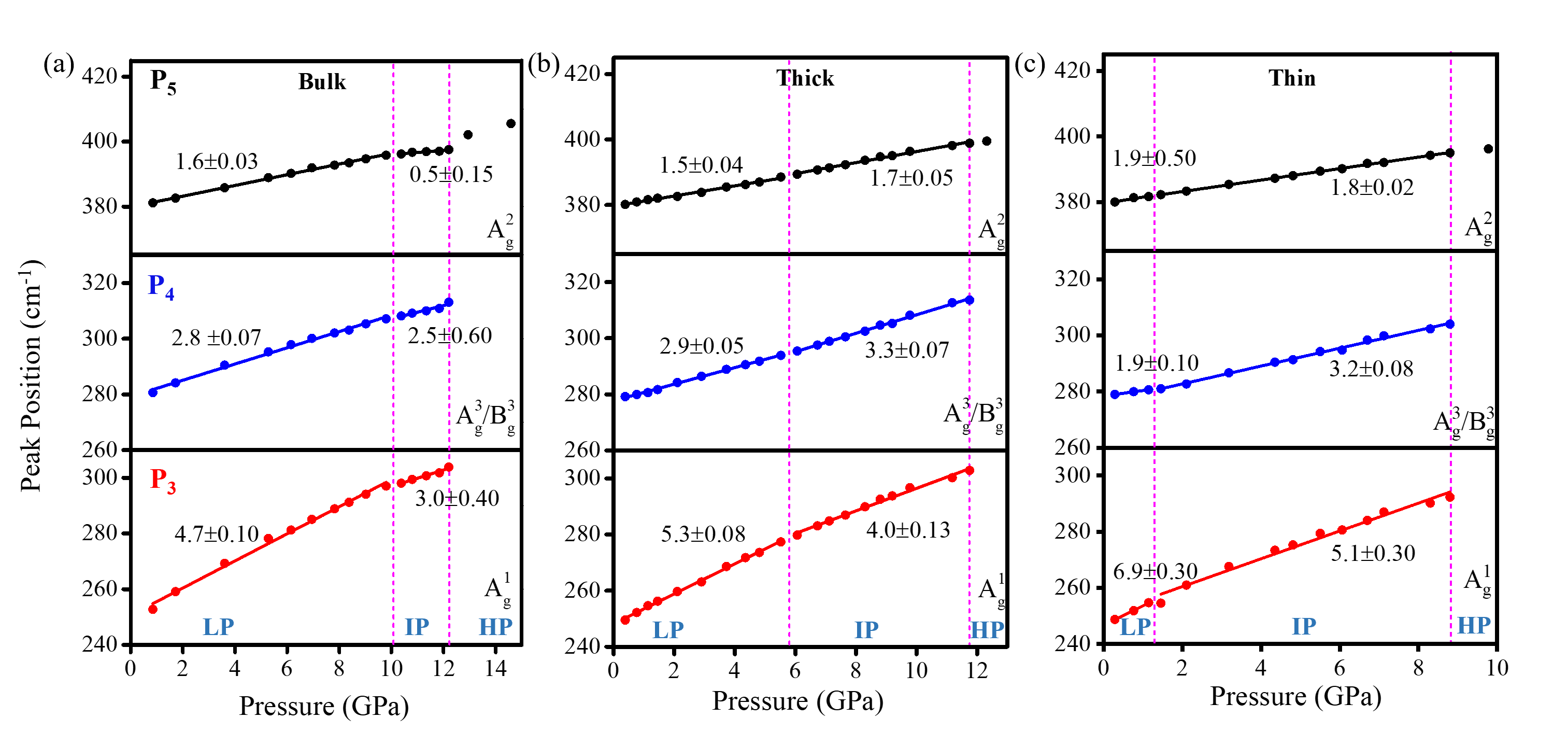}}
\caption{Pressure evolution of phonon modes with different layer numbers: (a), (b) and (c) shows the change in peak positions of P$_3$, P$_4$ and P$_5$ modes for bulk, thick and thin layered FePS$_3$ respectively. A critical pressure delineates the evolving trends in phonon modes across various layer thicknesses before the Raman modes disappear. DFT calculations along with experimental observations, determines the critical pressures and specifies the distinct pressure regions: low-pressure (LP), intermediate-pressure (IP) and high-pressure (HP) zone.  
\label{fig3}}
\end{figure}

To study the pressure dependence of the Raman modes of the three samples, we fit each mode of the Raman spectra using Lorentzian line shape function and plotted the peak positions of P$_3$, P$_4$ and P$_5$ modes with pressure in \fig{fig3}(a-c). 
The bulk, thick, and thin layer FePS$_3$ exhibit dissimilar trends in the pressure dependence of the Raman modes. Almost all the Raman modes disappear at different higher pressures. Before the modes disappear, a critical pressure separates the evolving trends of phonon modes for varying layer thicknesses. The DFT calculation is used to determine these critical pressures, and the results demonstrate a good match with the experimental observation, in which the phonon modes indeed exhibit a change in slope. Thus, three distinct regions - the low-pressure region (LP), the intermediate-pressure region (IP), and the high-pressure region (HP) can be identified within the overall pressure evolution. The modes exhibit a linear increase in all three regions and pressure dependent linear blue shifts for all three samples in LP and IP zones are tabulated in Table \ref{table:1}. The pressure dependent blue shift of the Raman modes is larger in thin layered FePS$_3$ than in the thicker sample. A similar trend is observed for graphene \cite{Proctor2009, Filintoglou2013} and TMDs \cite{Cheng2018} under pressure. For the thin layered sample, there are fewer vdW planes to buffer the pressure. Hence, more pronounced hardening of the vibrations and a higher rate of blue shift are observed.  The rate of increase of P$_3$  mode with pressure is higher than P$_4$  and P$_5$ mode for all the considered cases. The different pressure dependence of mode shifts can be understood by analysing the types of vibrations involved in these modes. The P$_3$ breathing mode involves out-of-plane vibration of sulphur atoms, whereas P$_5$ originates from symmetric stretching vibration of the P-S bonds and P$_4$ mode corresponds to in-plane stretching of P$_2$S$_6$ cluster \cite{Scagliotti1987, Bernasconi1988}. With increasing pressure, the compression of out-of-plane P$_3$ mode becomes more favourable than the in-plane P$_4$ and P-S bonds stretching P$_5$ mode. Further, in the case of bulk sample spectral separation between P$_5$ and P$_6$ increases with the pressure as it strengthens the out-of-plane P-P stretching P$_6$ vibration.

To rationalize the pressure-driven electronic phase transition as well as spin state transition, we have performed electronic structure calculations based on first-principles density functional theory (DFT) using the Vienna Ab Initio Simulation Package (VASP). The in-plane lattice parameters as well as interatomic distances decrease as pressure increases (Table S1).  In addition, the calculated magnetic moment of Fe atom (3.34 $\mu$B) at 0 GPa \cite{Lee2016, Zheng2019, Das2022} indicates that Fe$^{2+}$ ions in bulk FePS$_3$ are in a high spin state (S = 2) while the gradual increase of external pressure on FePS$_3$ leads to a spin-crossover from high spin state (HS) to low spin state (LS) where S = 0. This spin-state transition can be distinctly observed from the calculated magnetic moment value which decreases with the increase in pressure and finally becomes 0.00 $\mu$B/Fe-atom at the spin-state transition pressure of 10.80 GPa (Table S1).
The computational results indicate spin crossover at the onset of metallisation which is in line with the theoretical prediction by Zheng \textit{et al.}  \cite{Zheng2019}. Also, the x-ray emission spectroscopy study by Wang \textit{et al.} observed the collapse of Fe moment in the metallic phase suggesting spin crossover \cite{Wang2018}. On the other hand, Coak \textit{et al.}'s neutron powder diffraction experiment found persistent short-range magnetic orders that survive above room temperature where the Fe moment is similar to that at ambient pressure \cite{Coak2021}. However, in the current study, the SCO is solely based on DFT calculations which also predict insulator-to-metal transition at the same critical pressure with layer number dependency. Even though the computational study found cooperative behaviour of metallization and SCO, the current spectroscopic measurements cannot conclude the same. The explanation of these contradictory results with previous experimental studies is beyond the scope of the current experiment. In conclusion, this needs further investigation to resolve the unambiguousness of SCO in this type of system.

Bulk FePS$_3$  at 10.80 GPa undergoes a considerable decrease of in-plane lattice parameters (a = 5.65 $\AA$, b = 9.75 $\AA$), specially Fe-S distance indicating strain-induced lattice deformation and structural transition (Table S1).  
Density of states (DOS) analysis shn in Fig. S3(c) vividly exhibits that bulk FePS$_3$ at 0 GPa is an insulator with a band gap ($E_g$) of ~1.3 eV while the band gap narrows down with the increase of pressure and finally transforms into a metallic phase ($E_g$ = 0.00 eV) at 10.80 GPa indicating insulator to metal phase transition (as the PBE functional highly underestimates the band gap \cite{Perdew1983}, bandgaps were not determined from band structure calculation). The local and projected density of states (LDOS and PDOS) analysis further justifies the pressure-driven spin-crossover in bulk FePS$_3$. At P = 0 GPa, both the LDOS and PDOS of the Fe atom demonstrate the asymmetric nature of electronic density of states for up and down spin, whereas at transition pressure (10.80 GPa), the profile eventually becomes symmetric indicating spin pairing associated with the spin state transition (Fig. S3(d-f)).

Further, we have computed Raman spectra unveiling the optical phonon-property of bulk FePS$_3$ under pressure (Fig. S3(g)). Besides, the broadening and shifting of Raman peaks, two new peaks were generated at 326 and 342 cm$^{-1}$ due to structural distortion arising from the shift of Fe atoms in the hexagonal plane along with P atoms resulting change in the alignment of vdW planes at transition pressure (Fig. S3(h-i)). This structural change of bulk FePS$_3$, as well as the generation of new Raman peaks above 10 GPa, is in line with the earlier studies \cite{Lee2016, Haines2018, Das2022, Coak2021}. Hence, calculated Raman spectra further correlate the structural transition of bulk FePS$_3$ during spin state transition at 10.80 GPa. 

\begin{figure}[ht]
\centerline{\includegraphics[scale=0.8, clip]{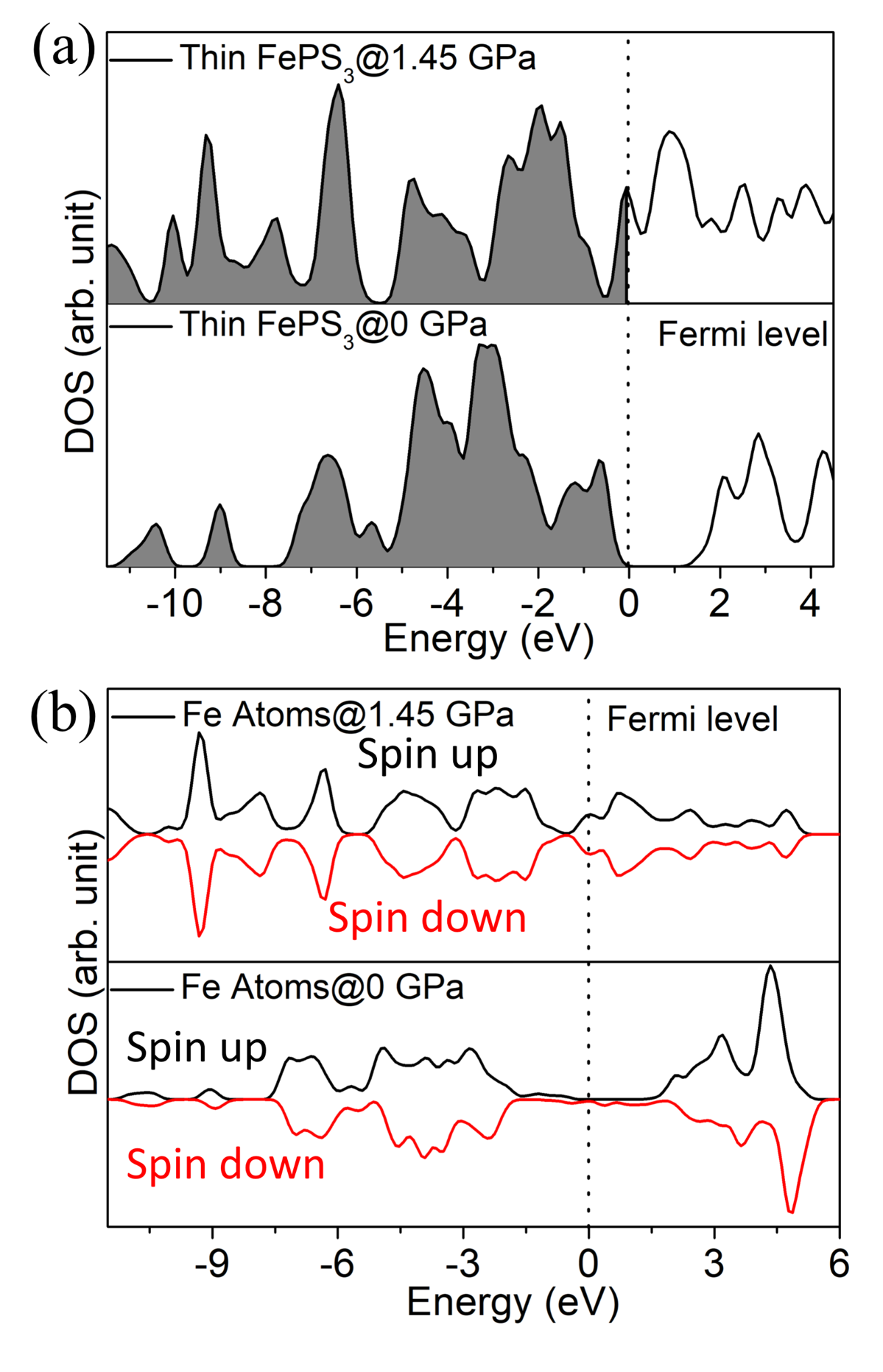}}
\caption{Density of States (DOS) calculations for FePS$_3$: (a) Total density of states (TDOS) and (b) Localized density of states (LDOS) of Fe atom for monolayer FePS$_3$ (thin sample). The asymmetric nature of the electronic density of states for up and down spin in LDOS eventually becomes symmetrical at a certain critical pressure, 1.45 GPa, where the bandgap reduces to zero, which indicates the transition of the spin state of Fe$^{2+}$ ion from S=2 to S=0. The critical pressure for both TDOS and LDOS studies increases with increasing layer numbers as shown in Fig. S4(c-d) for thick (6.06 GPa) and Fig. S3(c-d) for bulk sample (10.48 GPa).
\label{fig4}}
\end{figure}

\begin{figure}[ht]
\centerline{\includegraphics[scale=0.7, clip]{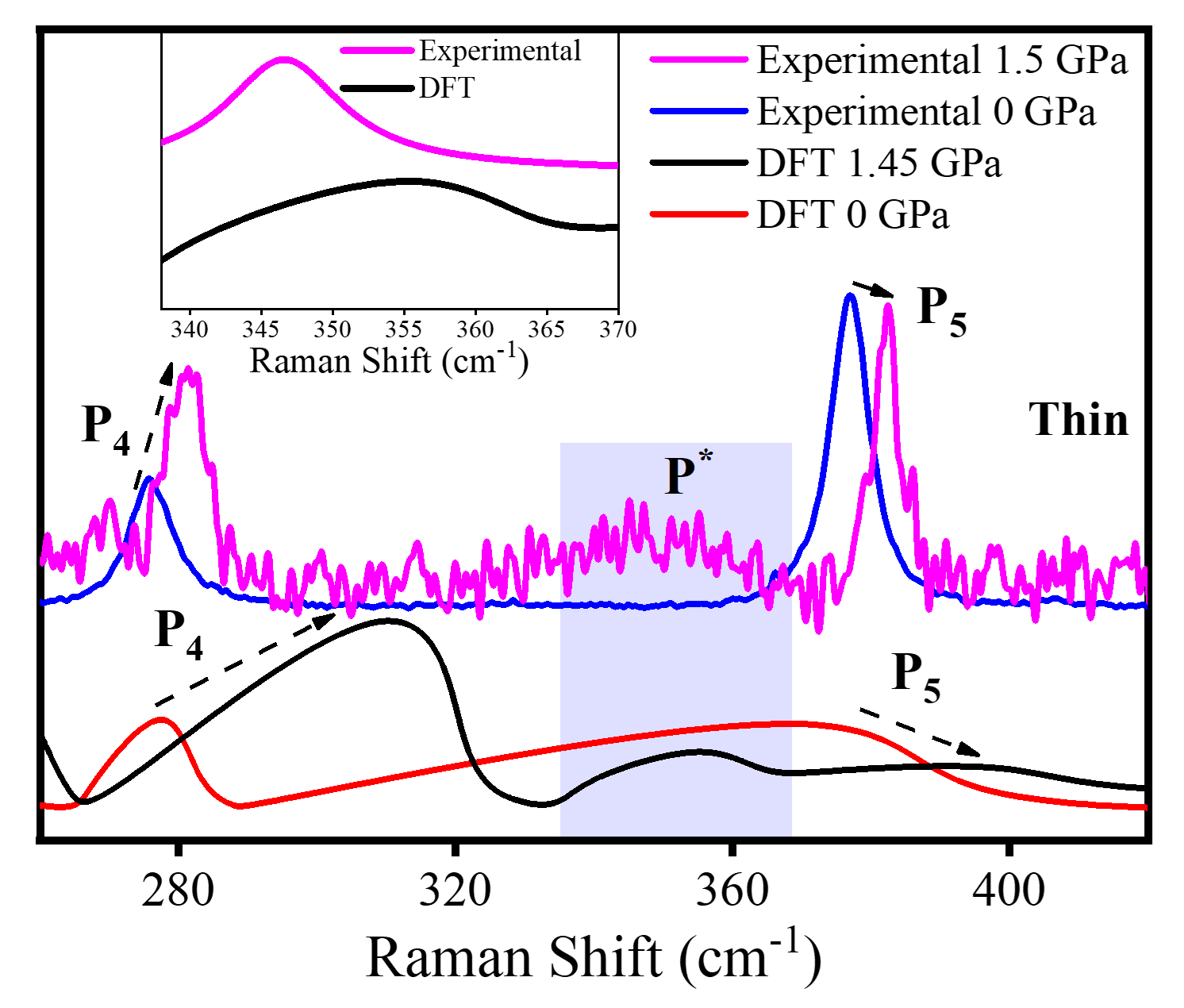}}
\caption{Experimental and computational Raman spectra for thin layer FePS$_3$: Dashed arrows highlight that both Raman spectra show a hardening of  P$_4$ and P$_5$ modes with an increase in pressure. The evolution of the new peak P$^\ast$ in the computational studies specified by the shaded region at the critical pressure can be mapped to the broad weak peak that evolves in the experimental high-pressure Raman spectra between P$_4$ and P$_5$. Inset is a comparative depiction of P$^\ast$ mode in the shaded region of the experimental (at 1.5 GPa) and computational spectra (at 1.45 GPa).
\label{fig5}}
\end{figure}

Similar pressure-induced lattice distortion, as well as structural transition, is also observed for the thick (Fig. S4(a-b)) and thin layer FePS$_3$ (Fig. S5(a-b)). The substantial change in in-plane lattice parameters and distance of Fe-S has similarly been observed here (Table S1). However the structural evolution is solely from DFT calculation and needs to be further analysed experimentally on exfoliated FePS$_3$ layers. For thick (Fig. S4(c)) and thin layer FePS$_3$ (\fig{fig4}(a)), DOS analysis demonstrates that the band gap narrows down with the increase of pressure and finally transforms into a metallic phase ($E_g$ = 0.00 eV) at 6.06 GPa and 1.45 GPa respectively indicating insulator to metal phase transition. Also from LDOS and PDOS analysis, unlike bulk, the thick (Fig. S4(d-f)) and thin layer FePS$_3$ (\fig{fig4}(b) and Fig. S5(d-f)) demonstrate spin state transition accompanying insulator-metal transition at a relatively lower external pressure of 6.06 GPa and 1.45 GPa respectively. Both the thick (Fig. S4(g)) and thin layer (Fig. S5(g)) FePS$_3$ exhibit similar Raman spectra under pressure except the higher frequency A$_g$ mode beyond 410 cm$^{-1}$ which is suppressed here in contrast to bulk FePS$_3$. 
A Raman mode P$^\ast$ evolves between 300 to 370 cm$^{-1}$ in the computational study for the bulk (Fig. S3(h-i)), thick (Fig. S4(h-i)) and thin (Fig. S5(h-i)) samples around the critical pressure at which the computed bandgap reduces to zero.  The calculated critical pressures for thin, thick and bulk samples match the onset of the IP region (\fig{fig3}) in experimental Raman spectra. The evolution of this new peak P$^\ast$ in the computational studies can be mapped to the broad weak peak that evolves in the experimental high-pressure Raman spectra between P$_4$ and P$_5$ for all three samples (Fig. S6). P$^\ast$ and the calculated magnetic moments for the three samples are represented in Fig. S7. Unlike in the bulk sample, the evolution of P$^\ast$ is more prominent in thin and thick samples due to less intense neighbouring phonon modes. \fig{fig5} depicts the comparative evolution of this broad phonon mode P$^\ast$ and the hardening of P$_4$ and P$_5$ modes in both experimental and computational Raman spectra for the thin flake. Similarly, for bulk and thick, see Fig. S6. P$^\ast$ may be a convolution of M4 to M8 modes observed in the high-pressure Raman study of FePS$_3$ in a thin crystalline pallet by Das \textit{et al.} \cite{Das2022} and was also supported by their theoretical calculation of Raman modes. The origin of this envelope requires further investigation.

\section{Discussion}
Using current computational studies and previous works on the same sample under external pressure, three distinct zones (LP, IP, HP) can be detected in pressure dependence of Raman modes (\fig{fig3}). The pressure-driven metallization and spin-crossover of the Fe$^{2+}$ ion from S = 2 ($t_{2g}^4e_g^2$)  to S = 0 ($t_{2g}^6e_g^0$) observed in the computational studies predicts an enhancement of \textit{d-p} metal-ligand hybridization which is almost at the onset of the experimental IP region. This indicates that the experimental IP region might be viewed as the beginning of the metallic phase which was previously described as a semiconducting phase by Zheng \textit{et al.} in their computational study \cite{Zheng2019}. Also, the IP region could be considered to be a bad metal, as predicted by Kim \textit{et al.},   where the IMT is mediated by $t_{2g}$ orbital of Fe$^{2+}$ ion, leaving $e_g$ gapped \cite{Kim2022}.  
Importantly, this IP region involves a structural change where weak vdW force favours slippage between layers \cite{pengchong}. Consequently, a sudden slippage among the layers takes place where the crystallographic  $c$-axis  becomes perpendicular to the $ab$-plane and Fe atoms in one layer are placed exactly above the adjacent layer \cite{Haines2018, Zheng2019}. It may be anticipated that this particular sliding will require higher pressure for a large number of layers under hydrostatic consideration. Because of this, as the number of layers decreases, the transition pressure for this initial structural phase transition that leads to the IMT will also decrease. 
On the other hand, at the beginning of the HP zone, an experimentally observed sudden volume collapse by Haines \textit{et al.} and Wang \textit{et al.} is accompanied by the creation of an in-plane Fe-Fe inter-metallic bonding, which delocalizes electrons \cite{Haines2018, Wang2018}. Thus, a complete metallization takes place in this regime followed by the Raman mode collapse in the micro Raman study.

To understand our results concerning the layer-dependent phase transition from the macroscopic viewpoint, any 2D stacking can be conceptualized as thin sheets experiencing anisotropic strain under hydrostatic pressure consideration (Fig. S8). In this configuration, each layer/sheet encounters the same amount of force ($F_t$) along the transverse direction. On the other hand, the force acting along the stairs will be equally distributed among all of the sheets (effective force, $F_{eff}$ = $\frac{F_l}{N}$, $N$ is the number of sheets). As a result, when the same amount of pressure is acting among all three directions, the out-of-plane strain will be thickness invariant, but the in-plane strain will be larger with fewer sheets since it is inversely proportional to the layer numbers.  Thus, it may be anticipated that any structural transition involving in-plane deformation will hold for a smaller number of sheets at a substantially lower external pressure. Interestingly, our experimental results justify the above argument if we consider the first transition point for the FePS$_3$ flakes which occurs at lower pressure for the thin layered samples (1.5 GPa for thin and 10.8 GPa for the bulk). Furthermore, around this pressure, if the effective in-plane strain is high enough for substantial orbital overlap, metallic behaviour may initiate in the sample. As a result, the transition from insulator to metal for thin layers will occur at a lower external pressure than that for thicker samples. Further, from the microscopic viewpoint, an electronic Hubbard model can be introduced by considering the spectral range (bandwidth $W$) as a function of layer number($N$) and pressure ($P$) with power-law scaling, which suggests IMT at relatively low pressure for a few-atomic layer flake (see Supplementary Information).

\section{Conclusions}
\label{sec:conclusion}

In conclusion, layer dependence of the critical pressure for the spin-crossover and the insulator-to-metal transition in a layered vdW magnet, FePS$_3$, has been systematically investigated via micro-Raman spectroscopy, supported by the first-principles DFT calculations. DFT calculations along with experimental observations determine the three distinct pressure regimes which reflect different trends in phonon modes across layer thicknesses. Detailed analysis of the computational results shows the SCO from a high (S = 2) to low spin state (S = 0) with a thickness-dependent critical pressure ($P_c$) which reduces to 1.5 GPa in 3-layer flakes compared to 10.8 GPa for the bulk counterpart. Though the disappearance of Raman modes in the experimental data indicates complete metallization at the high-pressure zone, computational studies suggest that IMT may be initiated with the metal-ligand charge transfer at much lower pressure. A phenomenological model explained that structurally anisotropic FePS$_3$ flake experiences more in-plane strain with decreasing layer numbers at a fixed pressure. The electronic Hubbard model where the layer number dependence in the form of power-law scaling, is introduced along with pressure in the bandwidth, suggests IMT at relatively low pressure for a few-atomic layer flake. The possibility of realizing 1-2 GPa in functional vdW materials with interfacial adhesion \cite{Trap2016} or adapting chemical pressure \cite{ChemicalPressure} may open up alternative strategies (see Supplementary Information) towards exploring novel phases in atomic-layer magnets for spintronic devices applications.


\begin{acknowledgements}

The authors would like to thank Goutam Dev Mukherjee, Sanjay Kumar and Anand Kumar for their technical help during the high-pressure experiments. The financial support (fellowships) from IACS, DST-INSPIRE and CSIR-UGC are greatly acknowledged. SD acknowledges the financial support from DST-SERB grant No. CRG/2021/004334. SD also acknowledges the technical research centre, IACS. AD thanks SERB grant CRG/2020/000301 for partial funding.

S.D. and B.M. conceived the project and designed the experiments. A.G. and S.M. prepared the samples and performed the initial characterization. B.M., S.D. and S.J. initiated the experiments at JNCASR taking inputs from C.N. Further, M.P., SDas, T.K. and S.M. carried out the pressure-dependent studies at IACS and confirmed the results. All DFT calculations were performed by R.J. with the inputs from A.D. SDas and S.D. developed the analytical model. All authors discussed the results and actively commented on the manuscript written by M.P., SDas, B.M., R.J. and S.D.

\end{acknowledgements}


\begin{table} [h!]
\caption{Rate of change of peak-shifts  for the phonon modes P$_3$, P$_4$ and P$_5$ with pressure for bulk, thick and thin samples:
The slopes (cm$^{-1}$/GPa) are calculated using linear fit for each sample and are mentioned for the low pressure (LP) and the intermediate pressure (IP) zones in the table. Pressure windows in two zones for the three samples are indicated in the parentheses in the GPa unit.}
    \centering
    \renewcommand{\thetable}{\arabic{table}}
    \begin{tabular} {c|ccc|ccc}
        \hline
       		Raman & & LP zone & & & IP zone &   \\

			modes & Bulk  & Thick & Thin & Bulk & Thick & Thin \\
			& (up to 10.8) & (up to 6.1) & (up to 1.5) & (10.8-12.9) & (6.1-12.3) & (1.5-8.9) \\
        \hline
	   P$_3$ & 4.7$\pm$0.1 		& 5.3$\pm$0.1  & 6.9$\pm$0.3 & 3.0$\pm$0.4 & 4.0$\pm$0.1 & 5.1$\pm$0.3\\        

	  P$_4$ & 2.8$\pm$0.1 & 2.9$\pm$0.1  & 1.9$\pm$0.1 & 2.5$\pm$0.6 & 3.3$\pm$0.1 & 3.2$\pm$0.1\\

	  P$_5$ & 1.6$\pm$0.1 & 1.5$\pm$0.1  & 1.9$\pm$0.5 & 0.5$\pm$0.2 & 1.7$\pm$0.1 & 1.8$\pm$0.1\\
       \hline 
    \end{tabular}

\label {table:1}
\end{table}


\end{document}


\title{Supplementary Information for Pressure induced insulator-to-metal transition in few-layer FePS$_3$ at 1.5 GPa}

\author{Bidyut Mallick$^{\dag}$}
\affiliation{Department of Applied Science and Humanities, Galgotias College of Engineering and Technology, Knowledge Park-II, Greater Noida, Uttar Pradesh - 201310, India}
\author{Mainak Palit$^{\dag}$}
\affiliation{School of Physical Sciences, Indian Association for the Cultivation of Science, 2A \& 2B Raja S. C. Mullick Road, Jadavpur, Kolkata - 700032, India
}

\author{Rajkumar Jana$^{\dag}$}
\affiliation{School of Chemical Sciences, Indian Association for the Cultivation of Science, 2A \& 2B Raja S. C. Mullick Road, Jadavpur, Kolkata - 700032, India
}
\author{Soumik Das}
\affiliation{School of Physical Sciences, Indian Association for the Cultivation of Science, 2A \& 2B Raja S. C. Mullick Road, Jadavpur, Kolkata - 700032, India
}

\author{Anudeepa Ghosh}
\affiliation{School of Physical Sciences, Indian Association for the Cultivation of Science, 2A \& 2B Raja S. C. Mullick Road, Jadavpur, Kolkata - 700032, India
}

\author{Janaky Sunil}
\affiliation{Chemistry and Physics of Materials Unit, Jawaharlal Nehru Centre for Advanced Scientific Research, Jakkur P.O., Bangalore - 560064, India
}
\author{Sujan Maity}
\affiliation{School of Physical Sciences, Indian Association for the Cultivation of Science, 2A \& 2B Raja S. C. Mullick Road, Jadavpur, Kolkata - 700032, India
}

\author{Bikash Das}
\affiliation{School of Physical Sciences, Indian Association for the Cultivation of Science, 2A \& 2B Raja S. C. Mullick Road, Jadavpur, Kolkata - 700032, India
}
\author{Tanima Kundu}
\affiliation{School of Physical Sciences, Indian Association for the Cultivation of Science, 2A \& 2B Raja S. C. Mullick Road, Jadavpur, Kolkata - 700032, India
}

\author{
Chandrabhas Narayana}
\affiliation{Chemistry and Physics of Materials Unit, Jawaharlal Nehru Centre for Advanced Scientific Research, Jakkur P.O., Bangalore - 560064, India
}

\author{Ayan Datta}
\affiliation{School of Chemical Sciences, Indian Association for the Cultivation of Science, 2A \& 2B Raja S. C. Mullick Road, Jadavpur, Kolkata - 700032, India
}
\author{Subhadeep Datta}
\email{sspsdd@iacs.res.in}
\affiliation{School of Physical Sciences, Indian Association for the Cultivation of Science, 2A \& 2B Raja S. C. Mullick Road, Jadavpur, Kolkata - 700032, India
}

\def\thefootnote{}\footnotetext{$^\dagger$ These authors contributed equally to this work}\def\thefootnote{\arabic{footnote}}

\maketitle

\section{Computational Details}

All the spin-polarized calculations were performed within the framework of density functional theory (DFT) using the plane-wave technique as implemented in the Vienna Ab Initio Simulation Package (VASP) \cite{Kresse1993}. The generalized gradient approximation method (GGA) parameterized by the Perdew-Burke-Ernzerhof (PBE) was used to account for the exchange-correlation energy \cite{Perdew1996}. DFT+$U$ method (Hubbard correction parameter) was used to account for the on-site coulomb repulsion and improve the description of localized Fe d-electrons in FePS$_3$ systems with $U_{eff}$ = 4.2 eV as recommended by the previous studies \cite{Lee2016}. The projector augmented wave potential (PAW) was used to treat the ion-electron interactions. The DFT-D2 empirical correction method proposed by Grimme was applied to describe the effect of van der Waals interactions \cite{Grimme2006}. Structural optimization at a specific external pressure was performed by adding the required pressure to the diagonals of the stress tensor. In all computations, the kinetic energy cut-off is set to be 520 eV in the plane-wave expansion. All the structures were fully relaxed (both lattice constant and atomic position) using the conjugated gradient method and the convergence threshold was set to be 10-4 eV in energy and 0.01 eV/$\AA$ in force. 
The Brillouin zone was sampled using a 5 $\times$ 5 $\times$ 5 Monkhorst-Pack k-point mesh for geometry optimization of bulk FePS$_3$ while a 5 $\times$ 5 $\times$ 1 Monkhorst-Pack grid was used for the thick and thin layered FePS$_3$ \cite{Lee2016} systems. A higher Monkhorst-Pack grid of 7 $\times$ 7 $\times$ 7 and 7 $\times$ 7 $\times$ 1 was used to calculate the electronic density of states (DOS) respectively for bulk and layered FePS$_3$ systems. The Raman intensity for each mode was calculated using the following equation \cite{Porezag1996}
\begin{equation}
    I_{Raman}=45\alpha'^2+7\beta'^2
\end{equation}
Where $\alpha'$ and $\beta'$ are the mean polarizability derivative and the anisotropy of the polarizability tensor derivative respectively and can be written as 
\begin{equation}
    \alpha'=\frac{1}{3}(\tilde{\alpha}'_{xx}+\tilde{\alpha}'_{yy}+\tilde{\alpha}'_{zz})
\end{equation}
\begin{equation}
   \beta'^2=\frac{1}{2}
   [(\tilde{\alpha}'_{xx}-\tilde{\alpha}'_{yy})^2
   +(\tilde{\alpha}'_{yy}-\tilde{\alpha}'_{zz})^2
   +(\tilde{\alpha}'_{zz}-\tilde{\alpha}'_{xx})^2
   +6(\tilde{\alpha}_{xy}'^{2}+
   \tilde{\alpha}_{yz}'^{2}+
   \tilde{\alpha}_{zx}'^{2})]
\end{equation}
Where  $\tilde{\alpha}'$ is the polarizability tensor and the respective derivatives are taken with respect to normal mode coordinate (Q), signifying the amount of displacement along one eigenvector of the system. The eigenvectors were obtained through direct diagonalization of the Hessian matrix. $\tilde{\alpha}'$ (per volume) was calculated with the VASP simulation package using density functional perturbation theory \cite{Kresse1993} by displacing the atoms along the eigenvectors of each mode twice. $\tilde{\alpha}'$  was approximated as a finite difference quotient and hence the Raman intensity of the mode was found. All these calculations were performed using a Python script \cite{Python2013} interfaced with VASP simulation code \cite{Kresse1993}.

Bulk FePS$_3$ was modelled considering the unit cell of monoclinic FePS$_3$ (space group C2/m) consisting of 20 atoms with lattice parameters a = 5.88 $\AA$, b = 10.19 $\AA$, c = 6.69 $\AA$. To model thick-layered FePS$_3$, we considered a three-layered periodic slab with 60 atoms while the thin-layered surface was modelled with the single-layered periodic slab with 20 atoms. In the case of thick and thin-layered FePS$_3$ surfaces, a vacuum layer of 20 $\AA$, was used in the direction perpendicular to the surfaces (along the Z-direction) to avoid spurious interactions between the slabs. Hence the values of $c$ parameter for monolayer and trilayer FePS$_3$ are increased disproportionately. Note that, the change in the bulk $c$ parameter is roughly -1.56\% for the given pressure region.

\section{Precision in pressure calculation from Ruby’s photoluminescence spectra:}

In the present study, we have calculated the pressure values from the calibration of Ruby’s photoluminescence spectra as reported by Shen \textit{et al.} \cite{Shen2020}. The temperature of the surroundings was kept fixed and a constant laser power was used throughout the experiment.  
For heat dissipation, two diamond anvils were attached with the Be-Cu sheet served as a good heat sink, although the samples were submerged in ethanol: methanol solution.

Interestingly, in the previous studies, there is a large dispersion in wavelength shift (0.0068 $< \delta \lambda$/$\delta T <$ 0.0076, in nm/K) of ruby’s fluorescence spectra varied with temperature reported by different authors.  However, in the calibration by Datchi \textit{et al.}  \cite{DATCHI2007} the leading order term of the wavelength shift in the fitted third order polynomial is $\delta\lambda_{R1}$ (296 $ < T <$ 900 K) $\sim$ 0.00746 $\delta T$, where $\delta\lambda_{R1}$($T$) = $\lambda_{R1}$($T$) − $\lambda_{R1}$ (296 K)  and R1 is the first peak of the photoluminescence doublet with $\lambda_{R1}$ (296 K) = 694.281 nm.  It is also mentioned that up to 600 K, it is well described by linear law i.e. $\delta\lambda_{R1}$ (296 $ < T <$ 600 K) = 0.00726(1) $\delta$T (Numbers in parentheses indicate the standard deviation of the fitting parameters).
Now, taking $\delta\lambda$/$\delta T$ $\sim$ 0.007 nm/K, we obtained $\delta$P/$\delta T$ $\sim$ 0.02 GPa/K.

The spectral resolution of our micro-Raman setup is 0.015 nm. To understand whether there is a change in ruby’s photoluminescence spectrum due to the heating effect (local heating or fluctuations in room temperature), a precision of order $\Delta\lambda$ $\sim$ 0.030 nm was considered. 
Therefore, any change in temperature of around 4 K ($\Delta\lambda$/($\delta\lambda$/$\delta T$) $\sim$ 4.3 K) cannot affect the fluorescence spectra. On the other hand, $\Delta\lambda$ $\sim$ 0.030 nm translates to a pressure change of order $\Delta P$ $\sim$ 0.08 GPa. In conclusion, the pressure values calculated from Ruby’s spectra are mentioned up to the first decimal place and the error is around $\pm$ 0.1GPa.

\section{Microscopic description of pressure-driven thickness-dependent IMT:}

We attempt to predict the insulator-to-metal transition by adopting the Hubbard model where the bandgap ($E_g$) between the upper and lower Hubbard subbands can be written as $E_g = U - W$ ($U$ is the onsite interaction and $W$ is the bandwidth). For strongly correlated Mott insulators, as in the case of FePS$_3$, we may consider $U \gg  W$ at ambient. For strongly correlated multi-electron $d^n$ system, $U$ is modulated by crystal-field splitting ($\Delta$), Hund's exchange coupling ($J$) and orbital degeneracy, and conventionally replaced by $U_{eff}$. Since the bandwidth increases with the reduction of inter-atomic spacing, metallic behaviour could be observed in the Mott insulator ($W \geq U_{eff}$) by lowering the inter-atomic distance. In our experiment, as the reduction of inter-atomic distance occurs due to external  Pressure ($P$), we may consider $W(P)=W(P_0)+c(P-P_0)^{\alpha}$, where $\alpha$ is a positive number, $c$ is a constant and $W(P_0)$ is the bandwidth at the ambient pressure ($P_0$). However, experimental data suggests that the transition pressure for IMT decreases with lowering the thickness \textit{i.e.} the layer number ($N$). Therefore, $W$ can be expressed as $W(N,P)=W(N)W(P)$, which may be assumed to scale as $N^{-\beta}P^{\alpha}$ ($\beta$ positive number). 
The behaviour of  $W(N, P)$ is different for the three samples and this affects the transition pressures identified in the Raman plots of Fig. 3, where $W(N, P)$ reaching a critical value results in gap-closure and ensuing metallization and collapse of Raman signatures.
When considering spin-crossover, the assumption that the Hubbard parameter is independent of pressure will be invalid. The effect of spin-crossover on the $U_{eff}$ in different $d^n$ configurations has different effects. Though $U_{eff}$ in the $d^5$ configuration decreases with increasing crystal field or pressure, the $d^6$ configuration exhibits a nonmonotonic increase due to the increase in correlation energy with the pressure. Whereas for $d^4$ and $d^7$, it is essentially pressure-independent \cite{dn2008}. On the other hand, the assumption of a monotonic increase in the crystal-field splitting will not be valid at any abrupt change in the lattice parameters. At these points, the $\Delta$($P$) value for isostructural transitions similarly undergoes a sharp change. Thus, depending on the $d^n$ configurations, these types of transitions might result in nearly immediate or delayed band closure. It requires more precise calculations for each case. We believe that further analytical model is required to improve the microscopic picture which might be useful to correlate the structural phase transition and metallization.

\section{Baseline correction and fitting of Raman spectra:}

 The experimentally obtained Raman spectra were processed further for  analysis and parameter extraction. The baseline was subtracted from the raw data followed by Lorentzian fitting of the individual peaks. \fig{figS9} shows Raman spectra of bulk, thick and thin FePS$_3$ with and without baseline correction. The fitting of individual modes of the experimental Raman spectrum of a bulk FePS$_3$ at ambient conditions is shown in  \fig{figS10}. \fig{figS11} shows the same for the experimental high-pressure Raman spectrum of a thin FePS$_3$ flake inside DAC. Similar processing of P$^{\ast}$ in the case of high-pressure experimental Raman spectra of bulk, thick and thin FePS$_3$ flake is explained in \fig{figS12}.

\section{Engineering interfaces and enclosures}

Nano-enclosures made from molecules/nanocrystals which are confined in the microscopic traps of layered magnets transferred onto conventional SiO$_2$/Si substrate can cause similar structural and conformational change observed on the flakes under hydrostatic pressure inside DAC (\fig{figS13}). Similarly local pressure up to 1-2 GPA can be realised in  heterostructures via interfacial addition (\fig{figS14}) and further with smaller nanoparticles trapped between them (\fig{figS15}).

\begin{figure}[h!]
\centerline{\includegraphics[scale=1, clip]{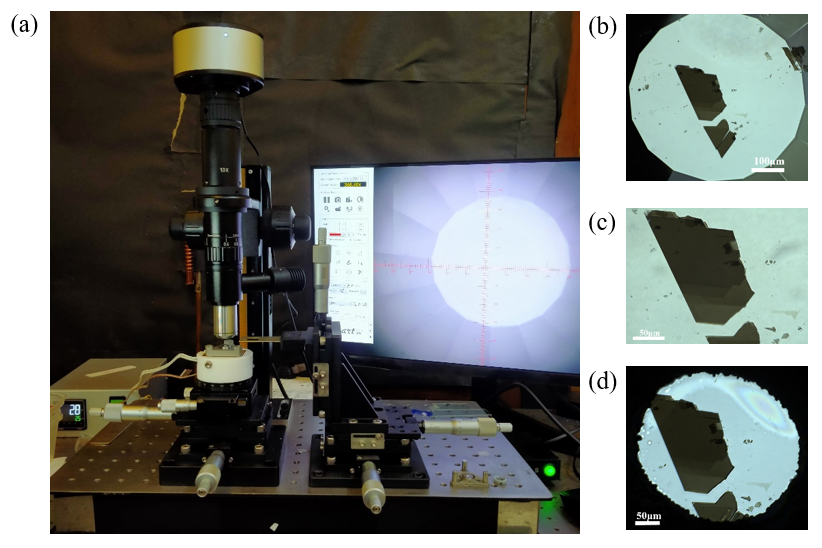}}
\renewcommand{\thefigure}{S\arabic{figure}}
\caption{Micromanipulator for targeted transfer of van der Waals flakes: (a) Exfoliated flakes were transferred onto the centre of the diamond culet using the micromanipulator under the attached optical microscope. The culet of the diamond anvil can be seen in the microscope display. (b) Optical micrograph of the exfoliated layered FePS$_3$ in transmission mode on diamond culet. (c) Zoomed images of the same region, where darker regions indicate more layer numbers. (d) Flakes in PTM after loading the gasket in the DAC.
\label{figS1}}
\end{figure}

\begin{figure}[h!]
\centerline{\includegraphics[scale=0.61, clip]{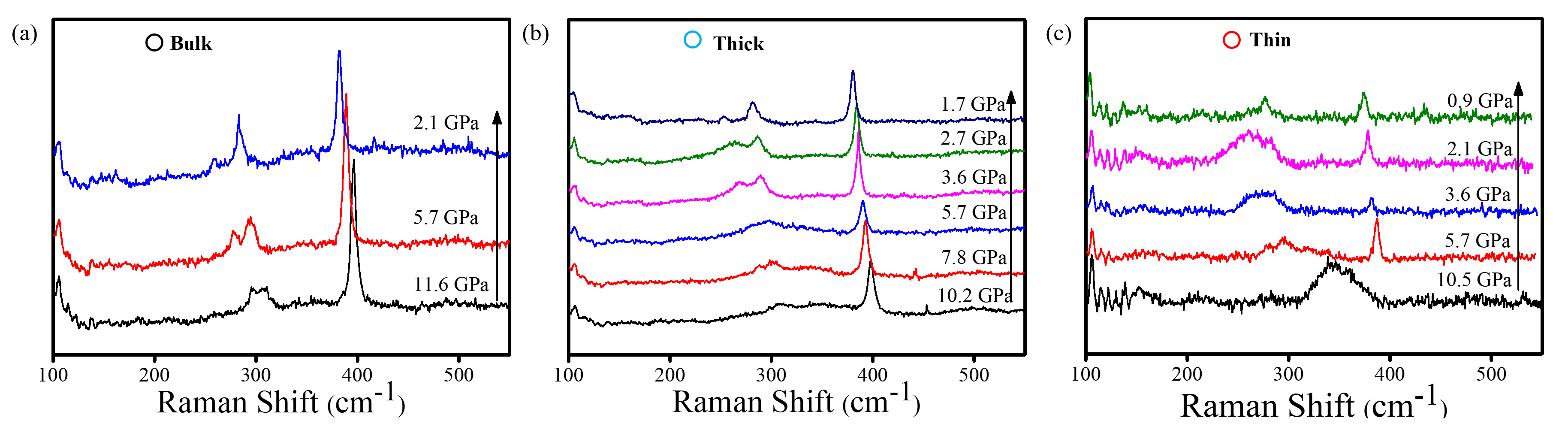}}
\renewcommand{\thefigure}{S\arabic{figure}}
\caption{Raman Spectroscopy of layered FePS$_3$ while releasing pressure: All the modes for (a) bulk, (b) thick and (c) thin FePS$_3$ flakes recover when pressure is released in the Diamond Anvil Cell. The arrow indicates the direction of release of pressure. 
\label{figS2}}
\end{figure}

\begin{table} [h!]
    \centering
    \renewcommand{\thetable}{S\arabic{table}}
    \caption{Computational study: Structural details and magnetic moment of bulk, thick and thin layer FePS$_3$ under external pressure.}
\label {table:1}
    \begin{tabular} {|c|c|c|c|c|c|c|c|}
        \hline
        FePS$_3$	& Pressure
	& In-plane lattice 	& Fe-Fe & Fe-S 	& P-P 	& Magnetic  \\

	& (GPa)	&   parameters ($\AA$)	& distance ($\AA$)	&  bond length ($\AA$)	&  bond length ($\AA$)	& Moment\\
	
	& 	&  	&	& 	&  	& ($\mu$B/Fe atom)\\
        \hline
        
        Bulk & 0 & a = 5.88 & 3.40 & 2.50	& 2.20	& 3.34 \\
         &  &  b = 10.19 &  & 	& 	&  \\      
        \hline
        & 3.60 & a = 5.73 & 3.30 & 2.41	& 2.18	& 2.75 \\
         &  &  b = 9.92 &  & 	& 	&  \\
        \hline
        & 5.28 & a = 5.67 & 3.28 & 2.40	& 2.17	& 2.69 \\
         &  &  b = 9.86 &  & 	& 	&  \\
        \hline
         & 10.80 & a = 5.65 & 3.26 & 2.24	& 2.17	& 0.00 \\
         &  &  b = 9.75 &  & 	& 	&  \\
        \hline
        Thick & 0 & a = 5.88 & 3.40 & 2.50	& 2.19	& 3.34 \\
         &  &  b = 10.19 &  & 	& 	&  \\
        \hline
         & 1.14 & a = 5.85 & 3.34 & 2.41	& 2.19	& 2.19 \\
         &  &  b = 9.98 &  & 	& 	&  \\
        \hline
        & 3.72 & a = 5.84 & 3.29 & 2.40	& 2.18	& 1.65 \\
         &  &  b = 9.91 &  & 	& 	&  \\
        \hline
         & 6.06 & a = 5.70 & 3.27 & 2.21	& 2.17	& 0.00 \\
         &  &  b = 9.83 &  & 	& 	&  \\
        \hline
        Thin & 0 & a = 5.88 & 3.41 & 2.51	& 2.20	& 3.35 \\
         &  &  b = 10.19 &  & 	& 	&  \\
        \hline
        & 0.25 & a = 5.84 & 3.39 & 2.34	& 2.19	& 0.83 \\
         &  &  b = 9.97 &  & 	& 	&  \\
        \hline
         & 0.50 & a = 5.78 & 3.31 & 2.31	& 2.15	& 0.06 \\
         &  &  b = 9.95 &  & 	& 	&  \\
        \hline
        & 1.45 & a = 5.72 & 3.30 & 2.28	& 2.14	& 0.00 \\
         &  &  b = 9.92 &  & 	& 	&  \\
        \hline
        
    \end{tabular}

\end{table}

\begin{figure}[ht]
\centerline{\includegraphics[scale=1, clip]{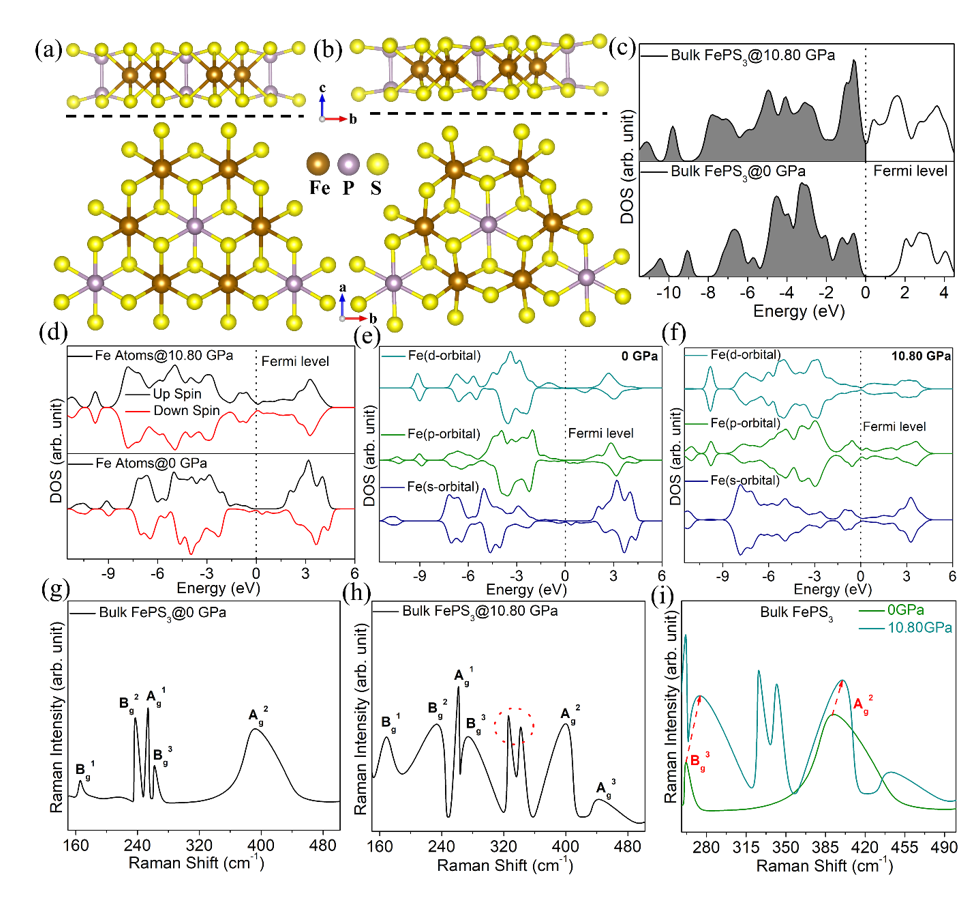}}
\renewcommand{\thefigure}{S\arabic{figure}}
\caption{Computational details of bulk FePS$_3$: Top and side views of the optimized structures of bulk FePS$_3$ at (a) P = 0 GPa, (b) P = 10.80 GPa, (c) Total density of states (TDOS), (d) Localized density of states (LDOS) of Fe atom, (e-f) Projected density of states (PDOS), (g-h) Simulated Raman spectra (red dotted encircled region indicates the generation of new peaks), (i) Blueshift in Raman Spectra of bulk FePS$_3$ at different pressures.
\label{figS3}}
\end{figure}

\begin{figure}[ht]
\centerline{\includegraphics[scale=1, clip]{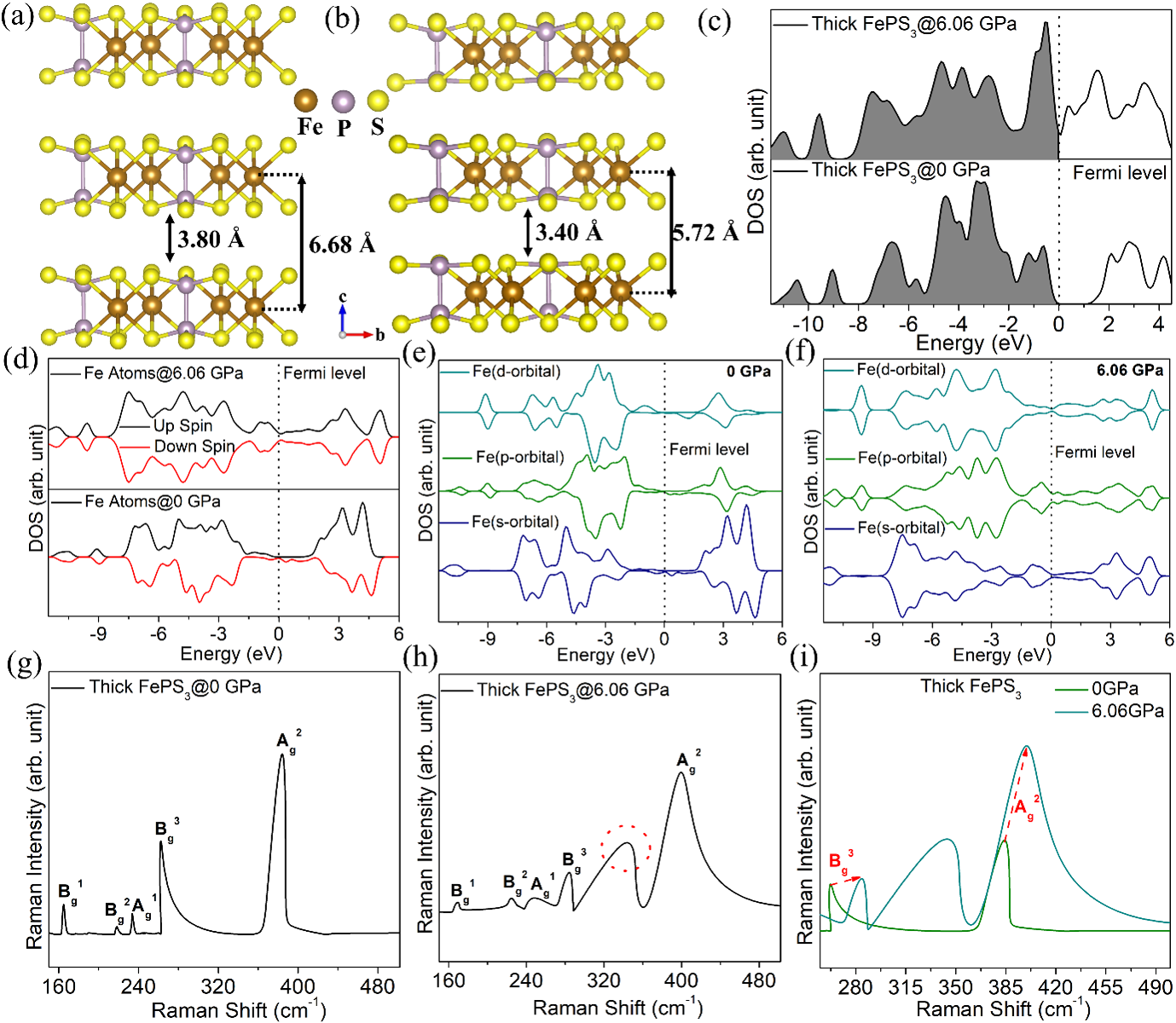}}
\renewcommand{\thefigure}{S\arabic{figure}}
\caption{Computational details of thick FePS$_3$: Optimized structures of thick layer FePS$_3$ at (a) P = 0 GPa, (b) P = 6.06 GPa, (c) Total density of states (TDOS), (d) Localized density of states (LDOS) of Fe atom, (e-f) Projected density of states (PDOS), (g-h) Simulated Raman spectra (red dotted encircled region indicates the generation of new peak), (i) Blueshift in Raman Spectra of thick layer FePS$_3$ at different pressures. 
\label{figS4}}
\end{figure}

\begin{figure}[ht]
\centerline{\includegraphics[scale=1, clip]{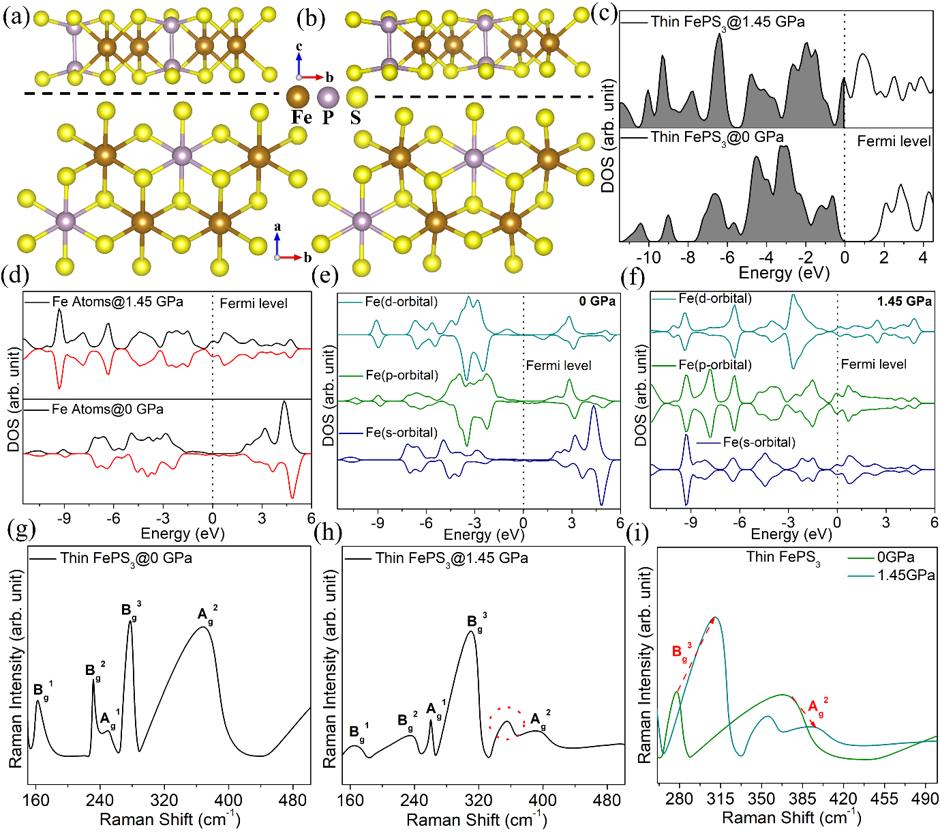}}
\renewcommand{\thefigure}{S\arabic{figure}}
\caption{Computational details of monolayer FePS$_3$: Top and side views of the optimized structures of thin layer FePS$_3$ at (a) P = 0 GPa, (b) P = 1.45 GPa, (c) Total density of states (TDOS), (d) Localized density of states (LDOS) of Fe atom, (e-f) Projected density of states (PDOS), (g-h) Simulated Raman spectra (red dotted encircled region indicates the generation of new peak), (i) Blueshift in Raman Spectra of thin layer FePS$_3$ at different pressures.
\label{figS5}}
\end{figure}

\begin{figure}[ht]
\centerline{\includegraphics[scale=0.621, clip]{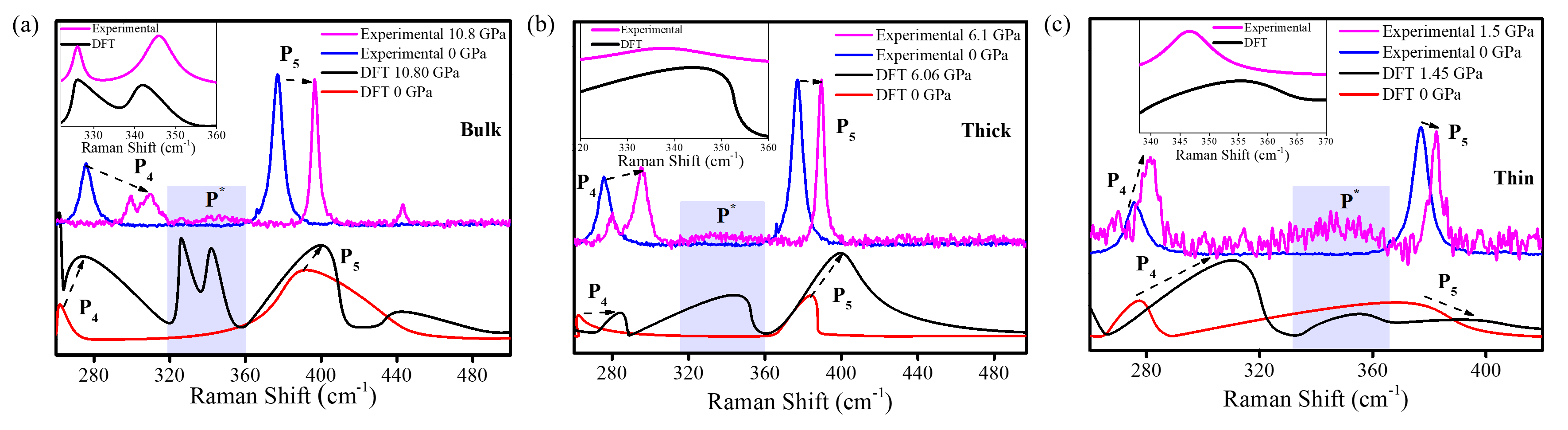}}
\renewcommand{\thefigure}{S\arabic{figure}}
\caption{Comparative depiction of experimental and computational Raman spectra of FePS$_3$: Dashed arrows highlight that both the Raman spectra for (a) bulk, (b) thick and (c) thin FePS$_3$ show a hardening of  P$_4$ and P$_5$ modes with an increase in pressure. The evolution of the new peak P$^\ast$ in the computational studies specified by the shaded region at the critical pressure can be mapped to the broad weak peak that evolves in the experimental high-pressure Raman spectra between P$_4$ and P$_5$ for all three samples. Inset in (a), (b) and (c) show P$^\ast$ mode of the respective experimental and computational spectra.
\label{figS6}}
\end{figure}

\begin{figure}[ht]
\centerline{\includegraphics[scale=0.57, clip]{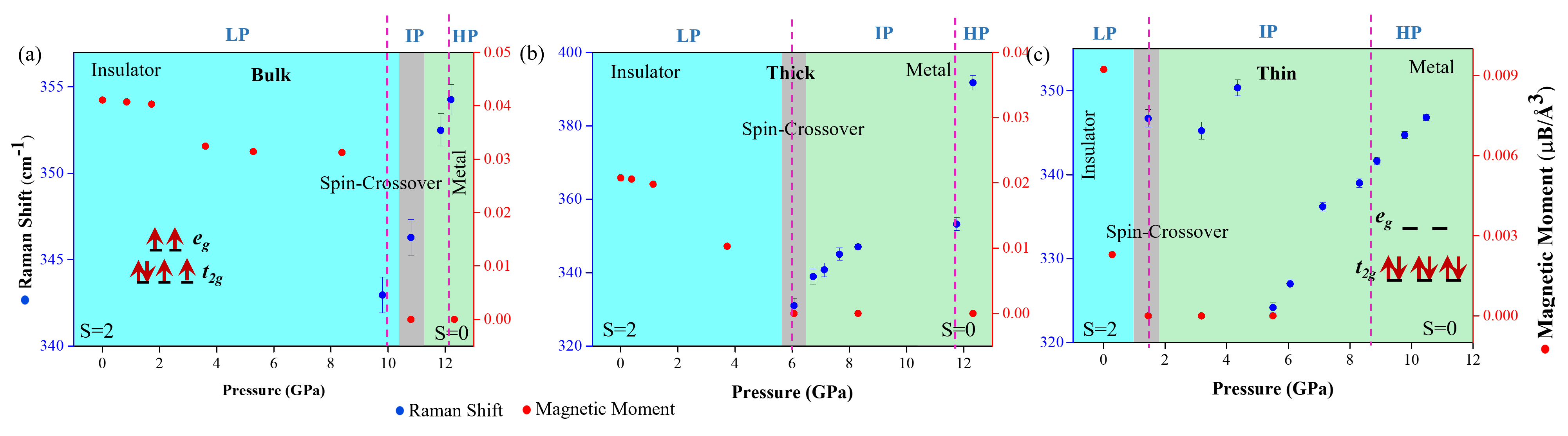}}
\renewcommand{\thefigure}{S\arabic{figure}}
\caption{Evolution of the P$^{\ast}$ mode and spin state transition from high spin (S=2) to low spin state (S=0): The blue dots depict the Raman shift of P$^{\ast}$ and the red dots depict the calculated magnetic moments for all three samples (a) bulk, (b) thick and (c) thin FePS$_3$. The sharp spin state transition in bulk samples becomes gradual with decreasing layer number. The critical spin crossover pressure decreases with decreasing layer number. The LP, IP and HP zones as described in Fig. 3 are indicated by broken lines for all three samples.
\label{figS7}}
\end{figure}

\begin{figure}[ht]
\centerline{\includegraphics[scale=0.5, clip]{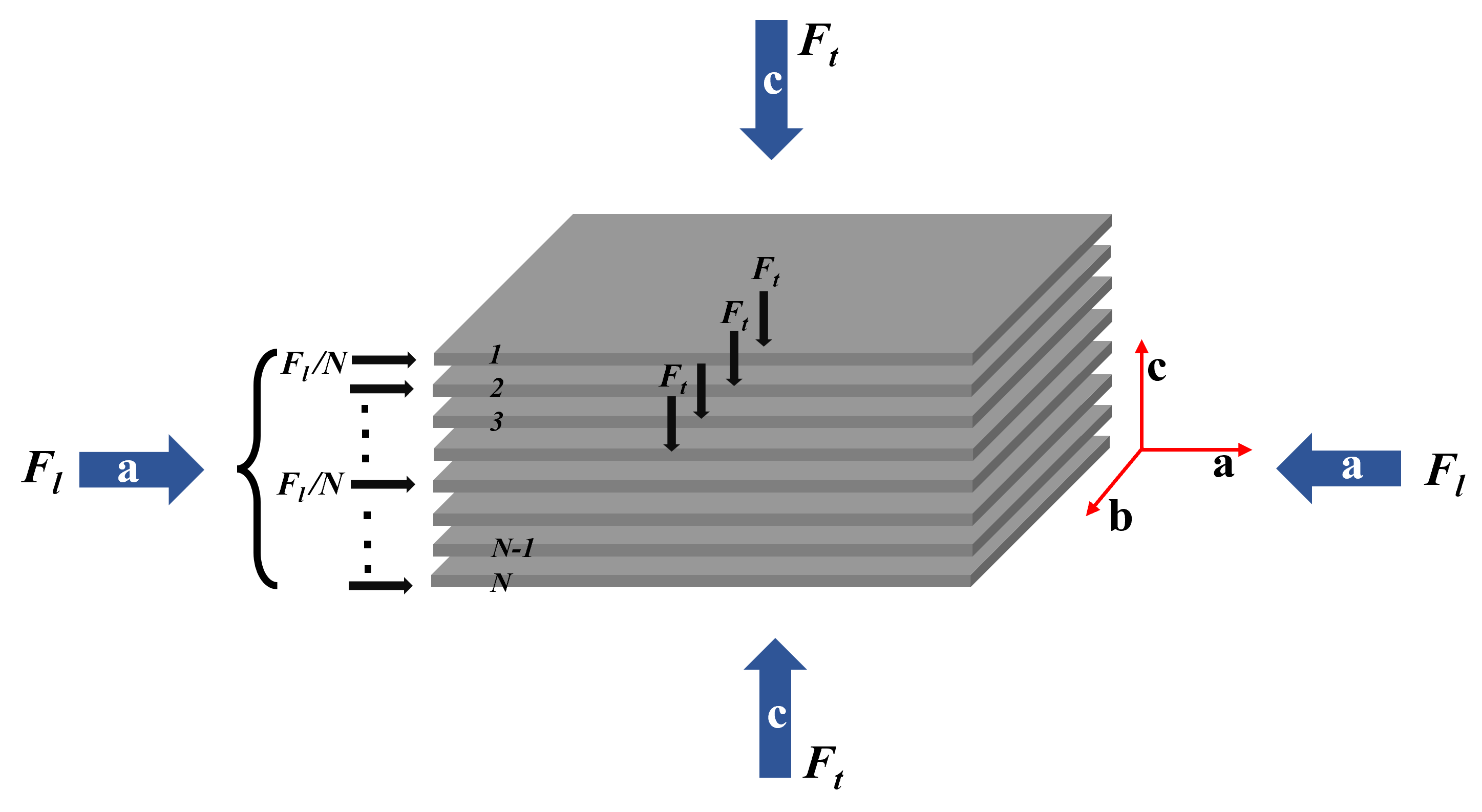}}
\renewcommand{\thefigure}{S\arabic{figure}}
\caption{Macroscopic model of 2D layered van der Waals material experiencing anisotropic strain under hydrostatic pressure: In this configuration, each sheet encounters the same amount of force ($F_t$) along the transverse direction (here, along the $c$ direction). The force acting along the stack (the $a$ or $b$ direction) will be equally distributed among all the sheets (effective force, $F_{eff}$ = $\frac{F_l}{N}$, $N$ is the number of sheets). The out-of-plane strain ($\frac{\delta l_c}{l_c}$) will be thickness invariant, but the in-plane strain  ($\frac{\delta l_a}{l_a}$ or $\frac{\delta l_b}{l_b}$) will be larger with fewer sheets since it is inversely proportional to the layer numbers. 
\label{figS8}}
\end{figure}

\begin{figure}[h!]
\centerline{\includegraphics[scale=0.55, clip]{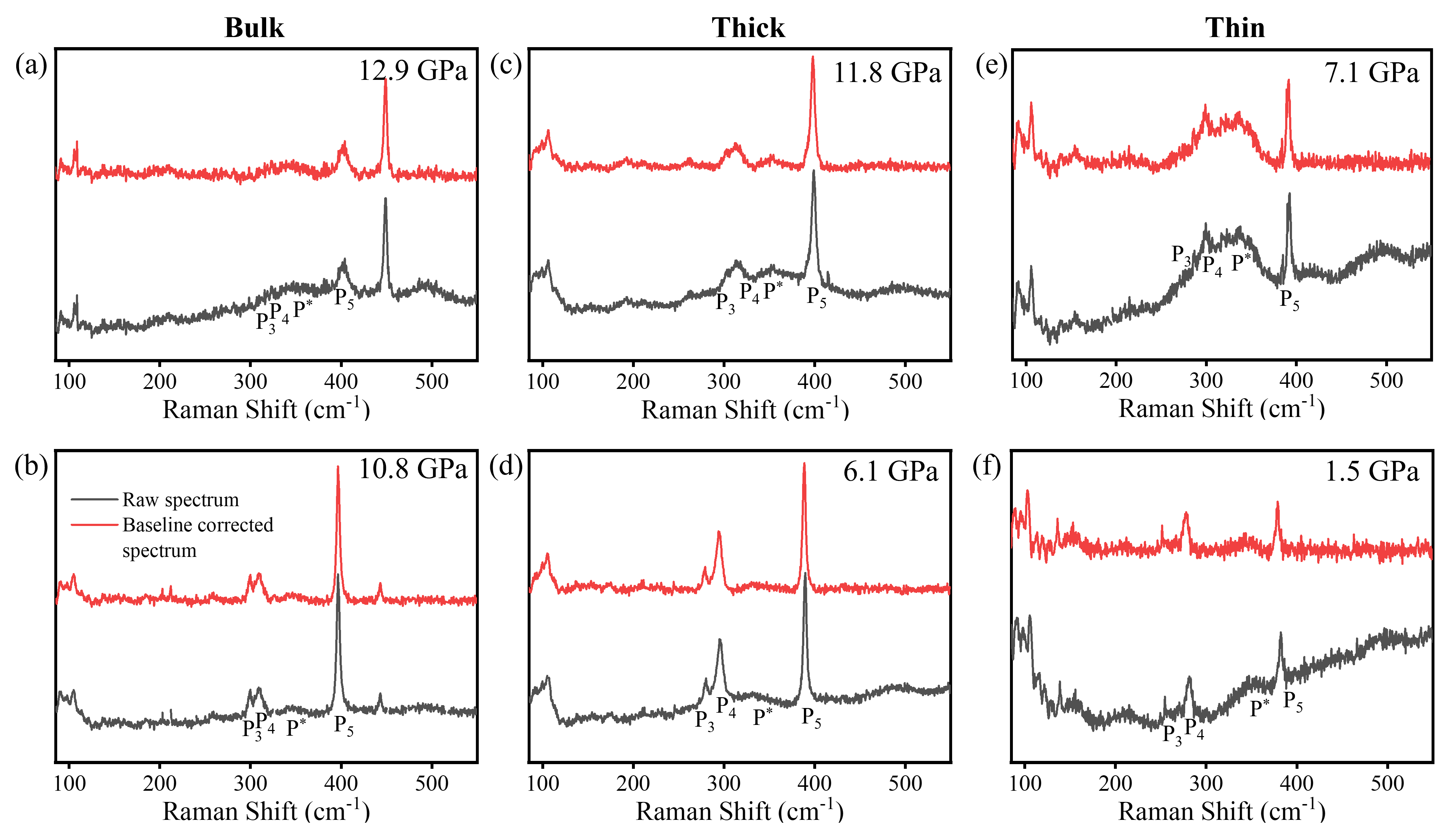}}
\renewcommand{\thefigure}{S\arabic{figure}}
\caption{Post-processing of experimental high-pressure Raman spectra of FePS$_3$ – baseline correction: Experimental (raw) and baseline corrected Raman spectra of FePS$_3$ are stacked in black and red lines respectively. (a) and (b) shows high-pressure Raman spectra of bulk FePS$_3$ at 12.8 GPa and 10.9 GPa. (c) and (d) shows high-pressure Raman spectra of thick FePS$_3$ flake at 11.8 GPa and 6.1 GPa. Finally, (e) and (f) shows high-pressure Raman spectra of a thin FePS$_3$ flake at 7.1 GPa and 1.5 GPa.  
\label{figS9}}
\end{figure}

\begin{figure}[h!]
\centerline{\includegraphics[scale=0.5, clip]{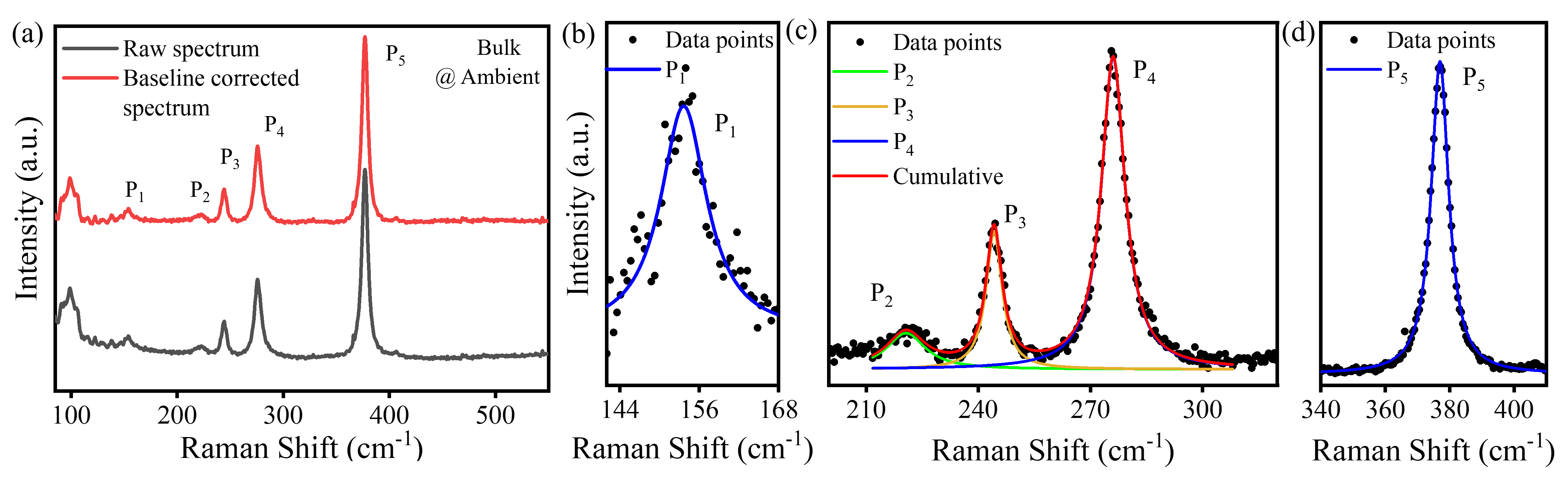}}
\renewcommand{\thefigure}{S\arabic{figure}}
\caption{Post-processing of the experimental Raman spectrum of a bulk FePS$_3$ at ambient condition – parameter extraction: (a) Experimental (black) and baseline corrected (red) Raman spectrum of bulk FePS$_3$. (b) shows the fitting of  P$_1$ using a Lorentzian function. (c) shows the fitting of  P$_2$, P$_3$ and P$_4$ using multiple peak fitting. The data points and the fitted functions are plotted with the cumulative (red), and (d) shows the fitting of  P$_5$.
\label{figS10}}
\end{figure}

\begin{figure}[h!]
\centerline{\includegraphics[scale=0.6, clip]{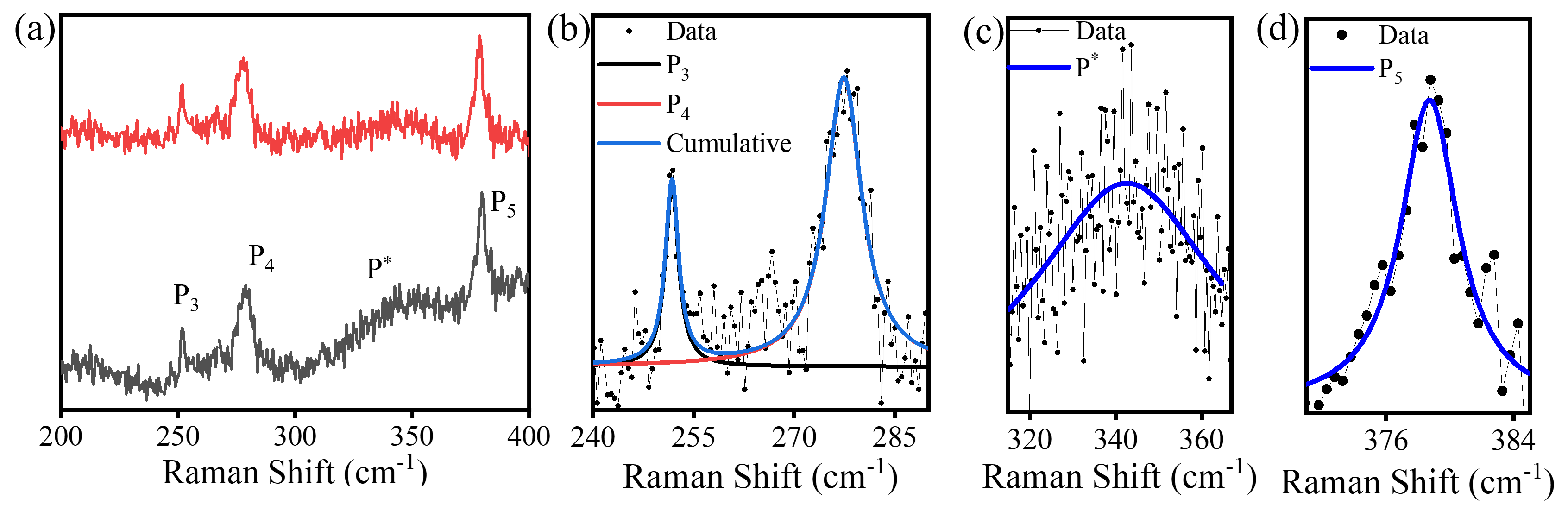}}
\renewcommand{\thefigure}{S\arabic{figure}}
\caption{Post-processing of the experimental high-pressure Raman spectrum of a thin FePS$_3$ flake inside DAC – parameter extraction: (a) Experimental (black) and baseline corrected (red) Raman spectrum of thin FePS$_3$ flake at 1.5 GPa. The Raman modes P$_3$, P$_4$, P$^{\ast}$ and P$_5$ are marked. (b) shows the fitting of  P$_3$ and P$_4$ using multiple peak fitting. The data points and the fitted function are plotted with the cumulative (blue). (c) shows the fitting of  P$^{\ast}$ using a Lorentzian function and (d) shows the fitting of  P$_5$.
\label{figS11}}
\end{figure}

\begin{figure}[h!]
\centerline{\includegraphics[scale=0.5, clip]{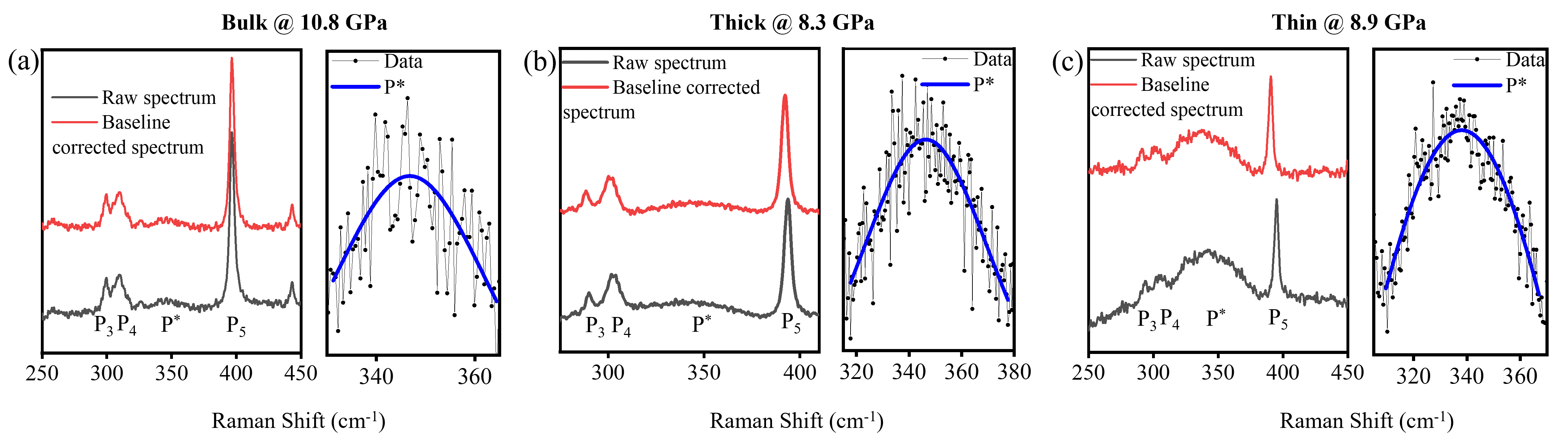}}
\renewcommand{\thefigure}{S\arabic{figure}}
\caption{Post processing of P$^{\ast}$ in three experimental Raman spectra of bulk, thick and thin FePS$_3$ flake at high-pressure – baseline correction and fitting of P$^{\ast}$: (a), (b) and (c) show experimental (black) and baseline corrected (red) Raman spectrum of bulk FePS$_3$ at 10.8 GPa, thick FePS$_3$ at 8.3 GPa and thin FePS$_3$ at 8.9 GPa respectively. The Raman modes P$_3$, P$_4$, P$^{\ast}$ and P$_5$ are marked the spectra. The P* is fitted using a Lorentzian function and plotted in blue.
\label{figS12}}
\end{figure}

\begin{figure}[h!]
\centerline{\includegraphics[scale=0.7, clip]{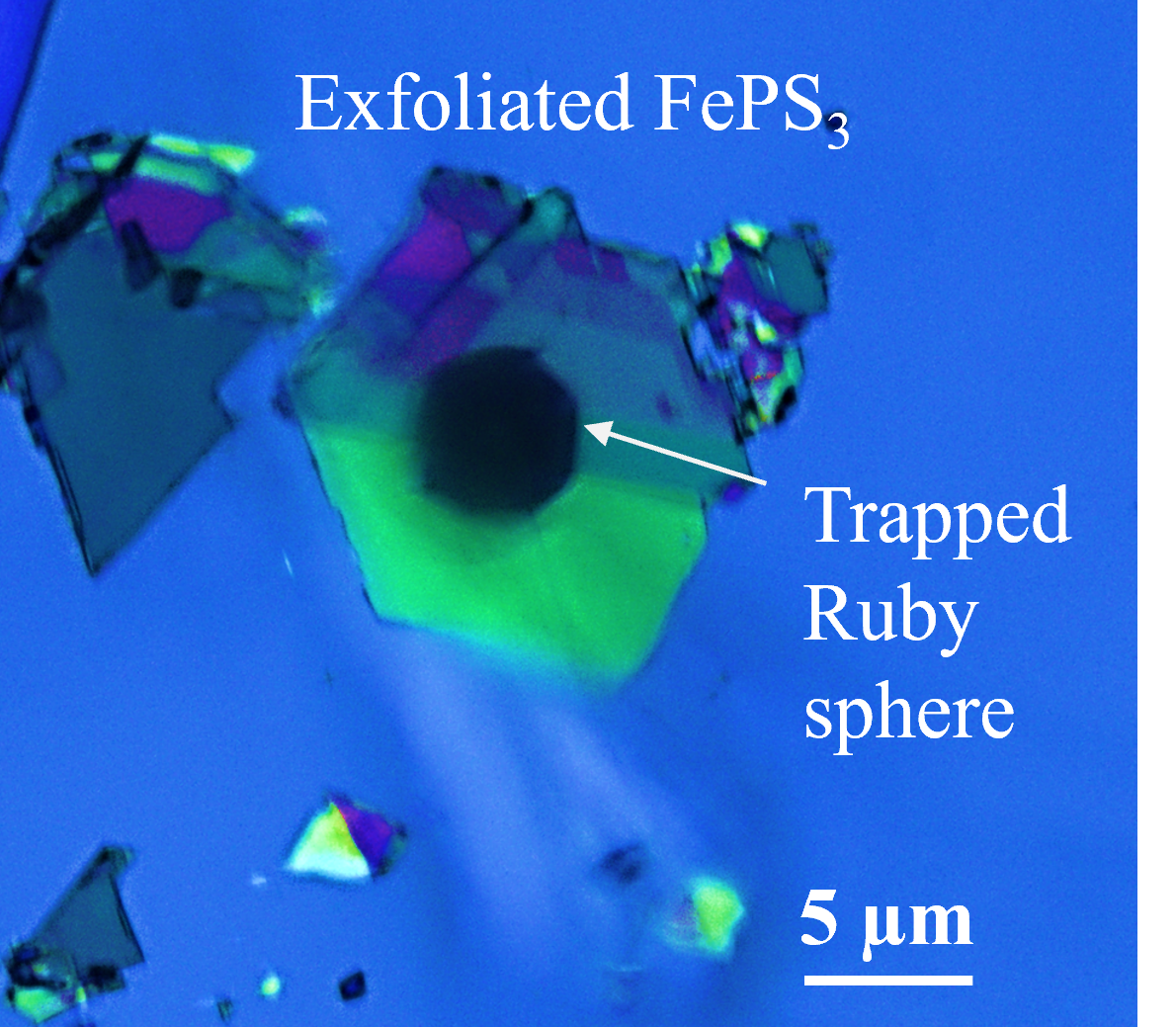}}
\renewcommand{\thefigure}{S\arabic{figure}}
\caption{Exfoliated FePS$_3$ flake strategically transferred on top of Ruby sphere (dark circular region) on SiO$_2$/Si wafer.
\label{figS13}}
\end{figure}

\begin{figure}[h!]
\centerline{\includegraphics[scale=0.6, clip]{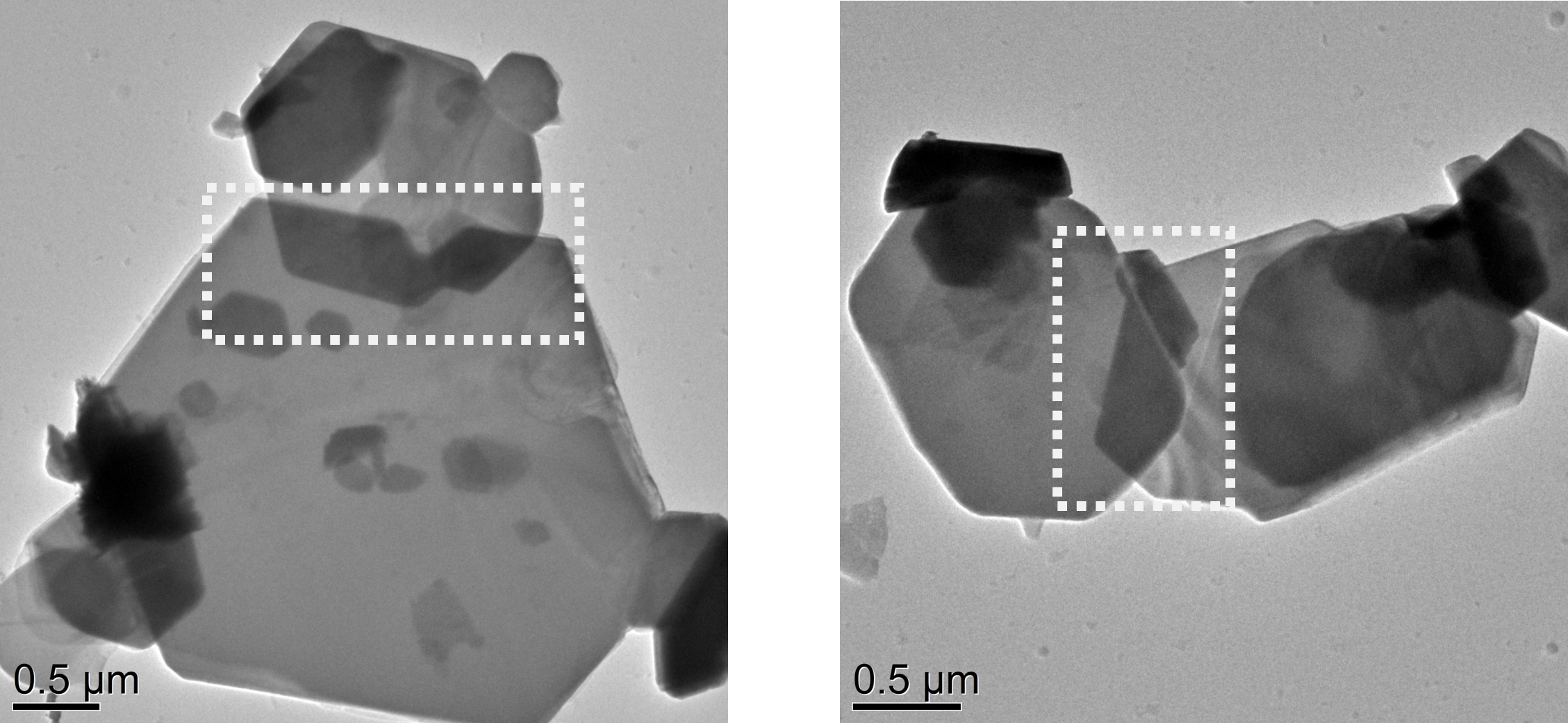}}
\renewcommand{\thefigure}{S\arabic{figure}}
\caption{Engineered heterostructures on top of TEM grid for probing the local strain at the edges of the heterostructure using the high-resolution TEM image in the overlap region. The overlap regions of interest are marked with rectangle with dashed border. 
\label{figS14}}
\end{figure}

\begin{figure}[h!]
\centerline{\includegraphics[scale=0.6, clip]{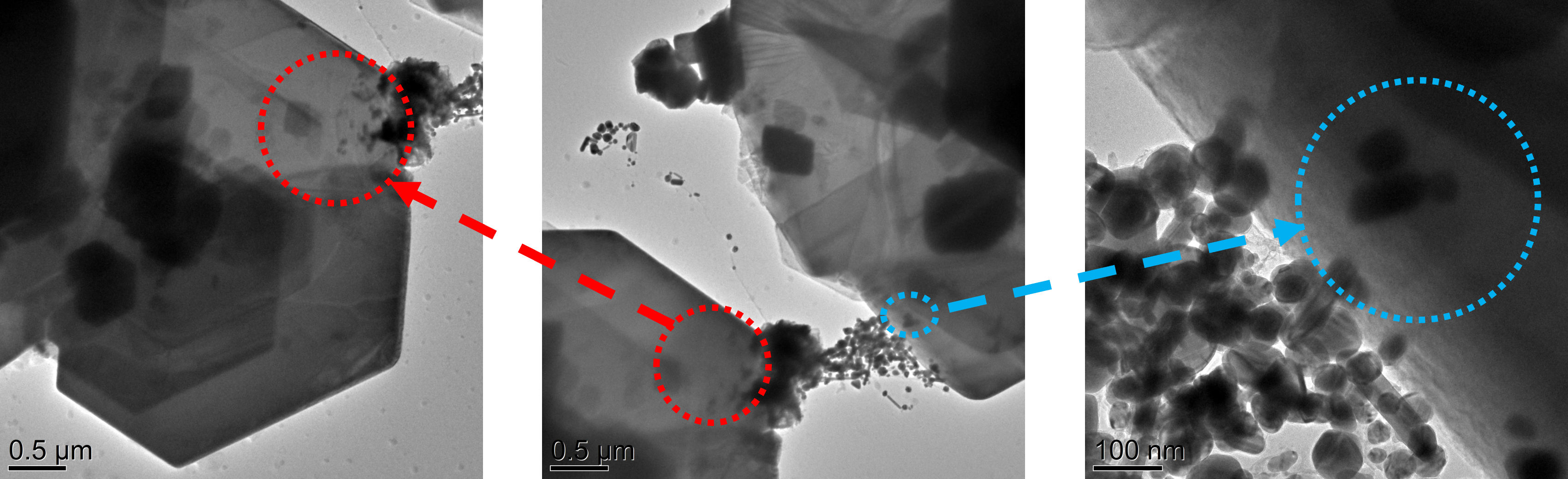}}
\renewcommand{\thefigure}{S\arabic{figure}}
\caption{Engineered FePS$_3$ heterostructures on top of TEM grid with Ag nanoparticles ($\sim$ 30 nm) sandwiched between them. The heterostructures are investigated using the high-resolution Transmission Electron Microscopy image in the overlap region. The overlap regions are marked with circles with dashed border. 
\label{figS15}}
\end{figure}
